\documentclass[prb,aps,twocolumn,amsmath,amssymb,floatfix,longbibliography]{revtex4-2}
\usepackage[breaklinks=true,colorlinks,citecolor=blue,linkcolor=blue,urlcolor=blue]{hyperref}
\usepackage{float}
\usepackage{bm}
\usepackage[normalem]{ulem}
\usepackage{epsfig,mathrsfs,color,latexsym,subfigure,marginnote,graphicx,verbatim,mathrsfs,color,array,amsmath,amsfonts,amssymb,graphicx}
\usepackage{tabularx}

\usepackage[utf8]{inputenc}
\usepackage{physics}

\begin{document}
\title{Linear magnetoresistance, anomalous Hall effect and de Haas-van Alphen oscillations in antiferromagnetic SmAg$_2$Ge$_2$ single crystals}

\author{Kanchan Bala}
\email{Contact author: kanchan.bala@tifr.res.in}
\affiliation{Department of Condensed Matter Physics and Materials Science, Tata Institute of Fundamental Research, Colaba, Mumbai 400005, India}

\author{Rahul Verma}
\affiliation{Department of Condensed Matter Physics and Materials Science, Tata Institute of Fundamental Research, Colaba, Mumbai 400005, India}

\author{Shovan Dan}
\affiliation{Department of Condensed Matter Physics and Materials Science, Tata Institute of Fundamental Research, Colaba, Mumbai 400005, India}

\author{Suman Nandi}
\affiliation{Department of Condensed Matter Physics and Materials Science, Tata Institute of Fundamental Research, Colaba, Mumbai 400005, India}

\author{Ruta Kulkarni}
\affiliation{Department of Condensed Matter Physics and Materials Science, Tata Institute of Fundamental Research, Colaba, Mumbai 400005, India}

\author{Bahadur Singh}
\affiliation{Department of Condensed Matter Physics and Materials Science, Tata Institute of Fundamental Research, Colaba, Mumbai 400005, India}

\author{A. Thamizhavel}
\email{Contact author: thamizh@tifr.res.in}
\affiliation{Department of Condensed Matter Physics and Materials Science, Tata Institute of Fundamental Research, Colaba, Mumbai 400005, India}

\begin{abstract}
Understanding the interplay among magnetism, electron correlations, and complex electronic structures in rare-earth materials requires both high-quality single crystals and systematic investigation of their electronic properties. In this study, we have successfully grown a single crystal of SmAg$_2$Ge$_2$ and investigated its anisotropic physical properties and de Haas-van Alphen (dHvA) quantum oscillations through experimental and theoretical approaches. SmAg$_2$Ge$_2$ crystallizes in the well known ThCr$_2$Si$_2$-type tetragonal structure with lattice parameters, $a~=~4.226$~\AA~ and $c~=~11.051$~\AA. Electrical transport and magnetization measurements indicate that it is metallic and exhibit antiferromagnetic ordering below the N\'{e}el temperature, $T_{\rm N}$ = 9.2~K. SmAg$_2$Ge$_2$ exhibits a linear non-saturating magnetoresistance, reaching $\sim 97$\% at $2$~K for applied magnetic field $B$~$\parallel~[001]$  and a significant anomalous Hall effect with an anomalous Hall angle of  $0.10-0.14$. Additionally, magnetization measurements reveal dHvA quantum oscillations for magnetic fields greater than $8$~T. Our calculated electronic structure, quantum oscillations, and anomalous Hall effect in the canted antiferromagnetic state closely align with experimental results, underscoring the role of complex electronic structure and spin-canting-driven non-zero Berry curvature in elucidating the physical properties of SmAg$_2$Ge$_2$.
	
\end{abstract}

\maketitle 
\section{INTRODUCTION}

Strongly correlated $f-$electron intermetallics constitute an exemplary class of materials with remarkable physical properties, which make them valuable for both fundamental research and technological applications~\cite{Buschow_1977,RevModPhys.81.1551,LaiSciAv2022,Ogunbunmi2021}. Previous studies have highlighted a range of diverse phenomena exhibited by these materials, including heavy-fermion superconductivity and hidden order in URu$_2$Si$_2$, unconventional triplet superconductivity in UTe$_2$, the Weyl-Kondo semimetal state in Ce$_3$Bi$_4$Pd$_3$, the topological Kondo insulator state in SmB$_6$, unconventional superconductivity and nontrivial electronic topology in CeRh$_2$As$_2$, and a Weyl semimetal phase exhibiting a giant anomalous Hall angle in TbPd(Pt)Bi, among other phenomena~\cite{HiddenOrder_URu2Si2,Superconductivity_URu2Si2,Ran2019_UTe2,Aoki2019_UTe2,WeylKondo,Neupane_KondoSmB6,Superconductivity_CeRh2As2,PRR_CeRh2As2,PRR_2021,PRB_2020,Zhu2023}. Such diverse properties in these materials stem from the competing strengths of various interactions, including Kondo effects, spin-orbit coupling, valence instabilities, and electrostatic crystal-field effects. Importantly, ternary rare-earth intermetallics \textit{RT$_2$X$_2$} (where \textit{R} represents a rare-earth element, \textit{T} is a transition metal, and \textit{X} is Si or Ge) exhibit complex magnetism, charge-instabilities, heavy-fermion behavior, superconductivity and nontrivial electronic structure owing to the presence of both $d$ and $f$ electron atoms~\cite{LaiSciAv2022,Nandi_2020,KHAN, LaiSciAv2022}. These materials generally adopt a ThCr$_2$Si$_2$-type body-centered tetragonal lattice with space group $I4/mmm$ (No. 139), where \textit{R}, \textit{T}, and \textit{X} atoms occupy Wyckoff positions $2a~(0,0,0)$, $4d~(0,0.5,0.25)$, and $4e~ (0,0,z)$, respectively, arranged along the tetragonal $c$-axis (see Fig.~\ref{fig1}). 

Among \textit{RT$_2$X$_2$} materials, SmMn$_2$Ge$_2$ is particularly interesting due to the presence of nearly pure $4f_{5/2}$ state of the Sm$^{3+}$ ion in its ground state~\cite{Neutron1995,PRB2004}. It exhibits various magnetic orderings, which are influenced by both the Sm and Mn sublattices. Notably, Mn atoms carry magnetic moments in addition to Sm atoms and Mn-Mn intralayer distance dictates the magnetic ground state at various temperatures~\cite{Neutron1995,PRB2004,Brabers1993}. SmMn$_2$Ge$_2$ exhibits multiple phase transitions as a function of temperature and pressure since the Mn-Mn intralayer distance resides close to a critical value of 2.87 \r{A}, which governs the different magnetic states in \textit{R}Mn$_2$Ge$_2$~\cite{PRB1994}. This material also displays a giant magnetoresistance and a temperature-tunable intrinsic anomalous Hall effect that varies from a substantial value at room temperature to zero at lower temperatures~\cite{Brabers1993,PRM_2024}. While the presence of nontrivial electronic structures and Hall responses makes SmMn$_2$Ge$_2$ appealing for the exploration of topological and nontopological properties, the coexistence of both Sm and Mn magnetic atoms complicates the understanding of their electronic and magnetic states and poses challenges in delineating the respective contributions of each magnetic atom to the overall magnetic state. These challenges could be more trackable if we replace one of the magnetic atoms with nonmagnetic metals. In particular, Au and Ag atoms, which possess completely filled $d$ states and are nonmagnetic, could serve as suitable replacements for Mn atom in Sm\textit{T$_2$}Ge$_2$. In this study, we successfully synthesized the single crystal of SmAg$_2$Ge$_2$ using the flux growth method and provided a detailed investigation of its magnetotransport and Fermi surface properties through various experimental and theoretical approaches. 
  
\section{Methods}

Single crystals of SmAg$_2$Ge$_2$ were grown using the self-flux technique, by taking advantage of the binary eutectic composition of ${\rm Ag:Ge}$, which melts at 650~$^{\circ}$C ~\cite {JOSHI}. The starting materials (3N-Sm, 5N-Ag, and 5N-Ge) were taken in a molar ratio of Sm:Ag:Ge = $1:16.25:6.75$. This mixture is placed in a high-quality alumina  (Al$_2$O$_3$) crucible and sealed in quartz ampoules under a partial pressure of Ar gas. The mixture was heated to 1150~$^{\circ}$C  at a rate of 60~$^{\circ}$C /h and maintained at this temperature for 24~h to achieve homogeneous melting, followed by slow cooling to 750~$^{\circ}$C at a rate of 2~$^{\circ}$C/h. Plate-like single crystals, with metallic lustre, were obtained by removing the excess flux by centrifugation. The phase purity of the grown crystal was checked by means of powder x-ray diffraction (XRD) in a PANalytical x-ray diffractometer equipped with a monochromatic Cu-$K_{\alpha}$ x-ray source ($\lambda = 1.5406$~\AA). Structural refinement of the XRD data was performed using Rietveld analysis with the FullProf software package \cite {RODRIGUEZCARVAJAL199355}. The crystals were oriented along the principal crystallographic directions using Laue diffraction and cut to the desired shape, using a wire electric discharge machine, for the anisotropic studies. The chemical composition of the crystals was confirmed through energy-dispersive x-ray spectroscopy (EDX). 

Magnetic measurements were performed in a SQUID magnetometer (MPMS, Quantum Design, USA). Electrical contacts were made using epoxy silver paste and gold wires of diameter 40~ $\mu$m on the crystal surface. Electrical resistivity, magnetoresistance, and Hall measurements were performed in a Quantum Design Physical Property Measurement System (PPMS, Quantum Design, USA). The magnetoresistivity and Hall resistivity data were obtained by symmetrizing and anti-symmetrizing the longitudinal and transverse Hall data measured under positive and negative magnetic fields, respectively.

Electronic structure calculations were performed within the framework of density functional theory using the projected augmented wave method as implemented in the Vienna ab-initio simulation package (VASP)\cite{Hohen,Bloch,Kresse1996,Kresse1999}. The generalized gradient approximation (GGA)~\cite{perdew1996generalized} was employed to account for exchange-correlation effects, and spin-orbit coupling was included self-consistently. The antiferromagnetic state of SmAg$_2$Ge$_2$ was modeled by treating the Sm $4f$-electrons as valence electrons, with an on-site Coulomb interaction added for Sm $4f$ within GGA+U scheme using U$_{eff}$ = 9 eV~\cite{HubbardU,LDAU}. Experimental lattice parameters were used for all calculations, and the internal atomic positions were relaxed until the residual forces on each atom were less than 10$^{-2}$~eV/\AA. A kinetic energy cut-off of 400 eV was employed for the plane-wave basis set. For bulk Brillouin zone sampling,  a $\Gamma-$centered 10$\times$10$\times$6 and 9$\times$9$\times$9 $k$-mesh were used for the conventional and primitive unit cells, respectively. The Fermi surface was generated by constructing a material-specific tight-binding Hamiltonian using the VASP2WANNIER90 interface~\cite{mostofi2008wannier90,Wtools}. Quantum oscillations were modeled by using the SKEAF code~\cite{skeaf}.

\section{RESULTS AND DISCUSSION}	

\subsection{ Crystal structure and magnetization of SmAg$_2$Ge$_2$}

\begin{figure}[ht!]
\includegraphics[width=1\columnwidth]{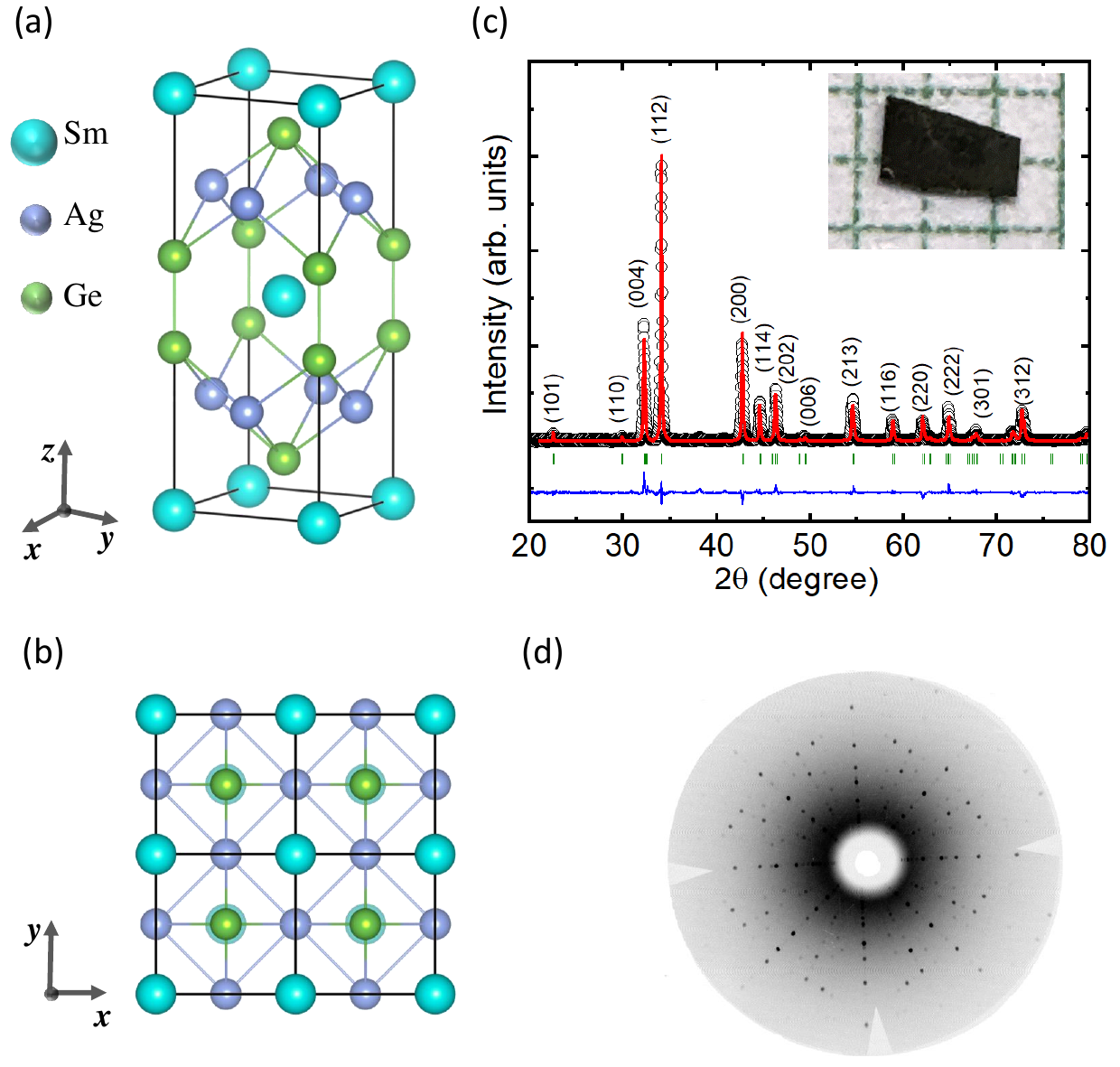}
\caption{(a) Crystal structure of SmAg$_2$Ge$_2$. (b) Top view of the crystal structure, highlighting the square net arrangement of Ag atoms. (c) Powder x-ray diffraction pattern with Rietveld refinement and (d) Laue diffraction pattern for (001) plane of SmAg$_2$Ge$_2$. The inset in (c) shows a typical single crystal image of SmAg$_2$Ge$_2$.}
\label{fig1}
\end{figure}

SmAg$_2$Ge$_2$ adopts a body-centered ThCr$_2$Si$_2$-type tetragonal crystal structure with space group $I4/mmm$. It features edge-connected AgGe$_4$ tetrahedra that stack along the tetragonal $c$-axis, mediated by Sm atoms (see Fig.~\ref{fig1}(a)). The Ag atoms form a two-dimensional (2D) planner square-net lattice with an Ag-Ag intralayer distance of 2.989~\AA~ (Fig.~\ref{fig1}(b)). Figure~\ref{fig1}(c) presents the room-temperature powder XRD pattern along with the Rietveld refinement. The XRD pattern and Rietveld analysis confirm reflections corresponding to the tetragonal structure of space group $I4/mmm$ with no detectable impurity peaks. During the refinement process, the thermal parameters of the atoms were fixed at zero, and atomic occupancy was maintained at stoichiometric values, as no significant improvement in the goodness fit was observed with variable occupancies. The extracted lattice parameters are $a~=~4.226$~\AA~ and $c~=~11.051$~\AA,  consistent with an earlier work~\cite {JOSHI}. Our analysis further indicates that Sm and Ag atoms occupy $2a$ and  $4d$ Wyckoff sites, respectively, whereas Ge atoms occupy the $4e$ Wyckoff position with a $z$ parameter of 0.3889. In Fig.~\ref{fig1}(d), we present the Laue diffraction pattern of SmAg$_2$Ge$_2$, obtained by exposing the flat plane of the crystal, which corresponds to the (001) plane. EDX analysis confirmed the composition close to the stoichiometric ratio of SmAg$_2$Ge$_2$.

\begin{figure}[t!]
\includegraphics[width = 0.49\textwidth]{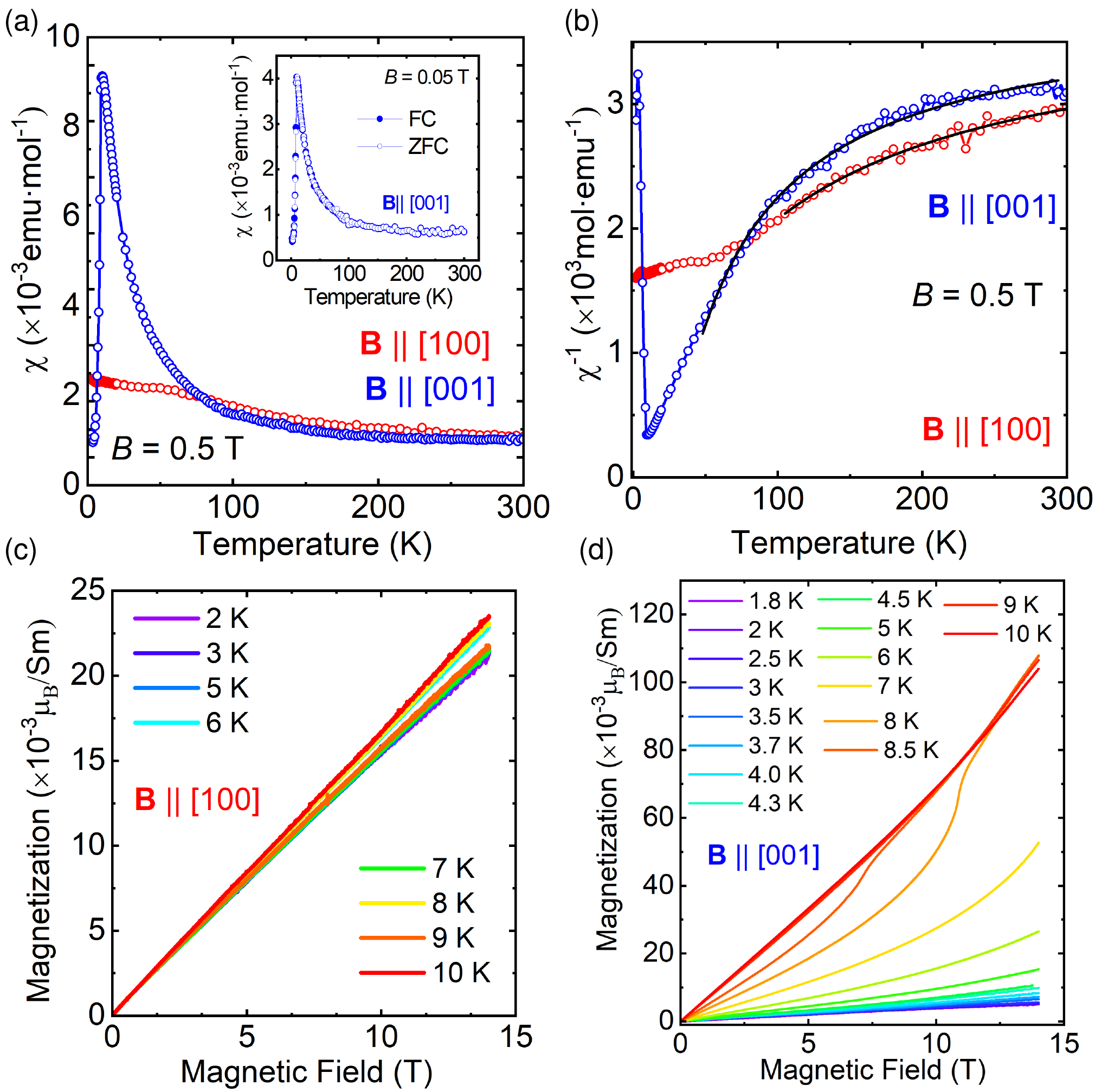}
\caption{(a) Magnetic susceptibility $(\chi)$ and (b) inverse magnetic susceptibility ($\chi^{-1}$) of SmAg$_2$Ge$_2$ as a function of temperature at in applied magnetic field $B= 0.2$~T. Data corresponding to $B$ $\parallel [100]$ and $B$ $\parallel [001]$ shown. Inset: Zero-field cooling (ZFC) and field cooling (FC) with an applied field $B$ = 0.05 ~T for $B$ $\parallel [001]$. Isothermal magnetization  of SmAg$_2$Ge$_2$ as a function of $B$ for (c) $B$ $\parallel [100]$ and (d) $B$ $\parallel [001]$.}
\label{fig2}
\end{figure}

Figure~\ref{fig2}(a) shows the magnetic susceptibility $\chi~(= M/B)$ of SmAg$_2$Ge$_2$ single crystal as a function of temperature under a static magnetic field of $B$ = 0.2 ~T, applied along the $a$-axis ($\chi_a$, $B$ $\parallel [100]$) and $c$-axis ($\chi_c$, $B$ $\parallel [001]$). The inset of Fig.~\ref{fig2}(a) presents the temperature dependence of magnetization under zero-field cooling (ZFC) and field cooling (FC) conditions with an applied field $B$ = 0.05 ~T along $c$-axis. At low temperatures, $\chi_c$ increases rapidly and exhibits a sharp drop at the N\'eel temperature $T_{\rm N}$ = 9.2~K, indicative of antiferromagnetic ordering. The magnetic susceptibility along the two principal crystallographic directions, $\chi_a$, and $\chi_c$ differ in magnitude, demonstrating significant magnetic anisotropy. Moreover, $\chi_c > \chi_a$ indicates that [001]-direction is the easy axis of magnetization. Figure~\ref{fig2}(b) presents the inverse magnetic susceptibilities $\chi_a^{-1}(T)$ and $\chi_c^{-1}(T)$ and the fits to Eq.~\ref{eq1} in the paramagnetic region (100~K $\leq$ T $\leq$ 300~K). The magnetic susceptibility of Sm compounds, excluding crystal field effects, can be described by:
\begin{equation}\label{eq1}
\chi = \frac{N_A \mu_{B}^2}{k_B} \left(  \frac{20}{7\Delta E} +  \frac{\mu_{eff}^2}{3(T-\Theta_p)} \right)
\end{equation}

where $\Delta E$ denotes the energy splitting between the excited state $J=\frac{7}{2}$ and the ground state $J=\frac{5}{2}$ of the Sm$^{3+}$ ion.  Fitting Eq.~\ref{eq1} to the paramagnetic region of the data (black curves in Fig.~\ref{fig2}(b)) yields $\mu_{\rm eff}$~=~0.45~$\mu_{\rm B}$/Sm, $\theta_{\rm p}$ = 22.86~K, and $\Delta E$ = 1968~K for $B$ $\parallel [001]$, and $\mu_{\rm B}$~=~0.58~$\mu_{\rm B}$/Sm, $\theta_{\rm p}~=~-3.75$~K, and $\Delta E$ = 1951~K for $B$ $\parallel [100]$. Comparing these results to the magnetic moment of the free Sm$^{3+}$ ion for the $J=\frac{5}{2}$ ground state of  $\mu_{\rm eff}$  = 0.85~$\mu_{\rm B}$/Sm, reveals deviations that indicate magnetic fluctuations within the system. Notably, the value of $\Delta E$ for $B$ $\parallel [001]$ is approximately 1500~K, similar to that of the free Sm$^{3+}$ ion. Nevertheless, the susceptibility clearly indicates antiferromagnetic ordering in SmAg$_2$Ge$_2$.

Figures~\ref{fig2}(c) and \ref{fig2}(d) illustrate the isothermal magnetization of SmAg$_2$Ge$_2$ as a function of the applied magnetic field $B~\parallel~[100]$ and $B~\parallel~[001]$, respectively. For $B~\parallel~[100]$, the magnetization exhibits a linear relationship for all applied fields and temperatures up to $10$~K, showing no signs of saturation or anomalies. However, oscillations in magnetization occur when the field exceeds $10$~T (see discussion below for details). In contrast, when $B~\parallel [001]~$, the magnetization displays a nearly linear dependence above the N\'{e}el temperature ($T_{\rm N}$~=~9.2~K). Below $T_{\rm N}$, significant non-linearity is evident, which indicates a possibility of spin flop or spin flip type transition at higher magnetic fields. Notably, despite the presence of metamagnetic transitions and sudden jump in magnetization, saturation of magnetization is not observed along this axis either up to a magnetic field of $14$~T. It is to be mentioned here that the metamagnetic transitions are discernible only for temperatures $5$~K and above, suggesting that the metamagnetic transitions appear at higher magnetic fields for temperatures less than $5$~K. Hence it is important to measure magnetization at high magnetic fields (greater than $14$~T).
 
\subsection{Heat capacity}

\begin{figure}[ht!]
\includegraphics[width=0.49\textwidth]{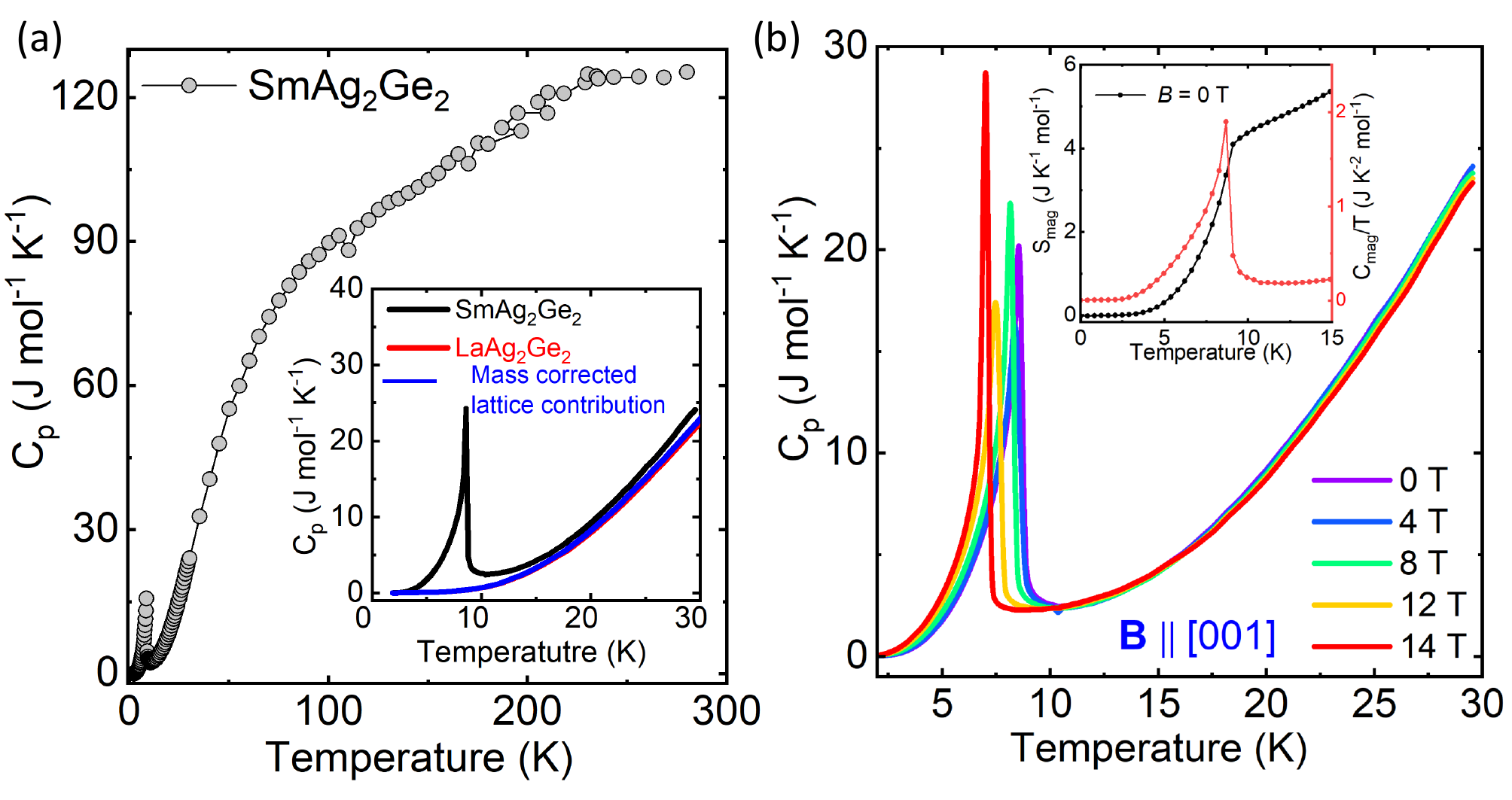}
\caption{(a) Temperature dependence of specific heat capacity ($C_{\rm p}(T)$) in the temperature range $2$-$300$~K, measured at zero magnetic field. Inset displays low-temperature $C_{\rm p}(T)$ for SmAg$_2$Ge$_2$, the reference nonmagnetic material LaAg$_2$Ge$_2$, and the estimated lattice contribution after correcting for the difference in formula masses of SmAg$_2$Ge$_2$ and LaAg$_2$Ge$_2$. (b) Specific heat capacity ($C_{\rm p}(T)$) as a function of temperature at different applied magnetic field. Inset: Magnetic contribution to the heat capacity ($C_{\rm mag}$/T) (right axis) and entropy ($S_{\rm mag}(T)$) (left axis) as a function of temperature for SmAg$_2$Ge$_2$ at zero magnetic field.}
\label{fig3}
\end{figure}

The heat capacity of SmAg$_2$Ge$_2$ single crystal as a function of temperature is shown in Fig.~\ref{fig3}(a). A pronounced $\lambda$-type anomaly is observed at $9.2$~K (inset of Fig.~\ref{fig3}(a)), confirming the bulk magnetic ordering. Additionally, $C_{\rm p}(T)$ attains a value of $\approx$122~J mol$^{-1}$K$^{-1}$ at $300$~K, which is close to the expected classical Dulong-Petit value of = $3nR$ =124~J mol$^{-1}$K$^{-1}$, where $n=5$ is the number of atoms per formula unit cell and $R$ is the molar gas constant~\cite {Anand2014}.

To resolve the magnetic contribution to the specific heat capacity of SmAg$_2$Ge$_2$, we compare ($C_{\rm p}(T)$) with the specific heat capacity of LaAg$_2$Ge$_2$, along with the mass-corrected lattice contribution (see inset of Fig.~\ref{fig3}(a)).  The transition temperature $T_{\rm N}$ decreases with increasing magnetic field, indicating the long-range antiferromagnetic nature of the ordering [see Fig.~\ref{fig3}(b)]. Notably, long-range interactions in antiferromagnetic materials typically compete with the applied magnetic field, reducing the ordering temperature. As evidenced by the magnetization data, Sm $4f$ moments tend to align along the direction of the applied field for $B$ $\parallel [001]$. Consequently, the transition temperature shifts to lower values with increasing field strength, consistent with our magnetic measurements

We plot magnetic contribution to the heat capacity, $C_{\rm mag}$ of SmAg$_2$Ge$_2$, which is obtained by subtracting the mass-corrected specific heat of LaAg$_2$Ge$_2$ from ($C_{\rm p}(T)$) at zero applied magnetic field in the inset of Fig.~\ref{fig3}(b). We further estimate the magnetic entropy $S_{\rm mag}(T)$ by integrating $C_{\rm mag}/T$ versus temperature data~\cite{PhysRevB.93.094422}. The temperature dependence of magnetic entropy $S_{\rm mag}(T)$ is shown the inset of  Fig.~\ref{fig3}(b). $S_{\rm mag}(T)$ tends to saturate to a value of $\approx 5.8~$Jmol$^{-1}$K$^{-1}$, which is $\sim42 \%$ of the expected theoretical entropy of $R \ln(2J + 1)$ with $J = 5/2$ for Sm$^{3+}$ ion, where $R$ is the universal gas constant. This value, however, close to $R \ln(2)$ with $J = 1/2$, suggesting that the ground state is a doublet due to the low crystal symmetry of SmAg$_2$Ge$_2$.

\subsection{Electrical resistivity and magnetoresistance}

\begin{figure}[t!]
\includegraphics[width=0.49\textwidth]{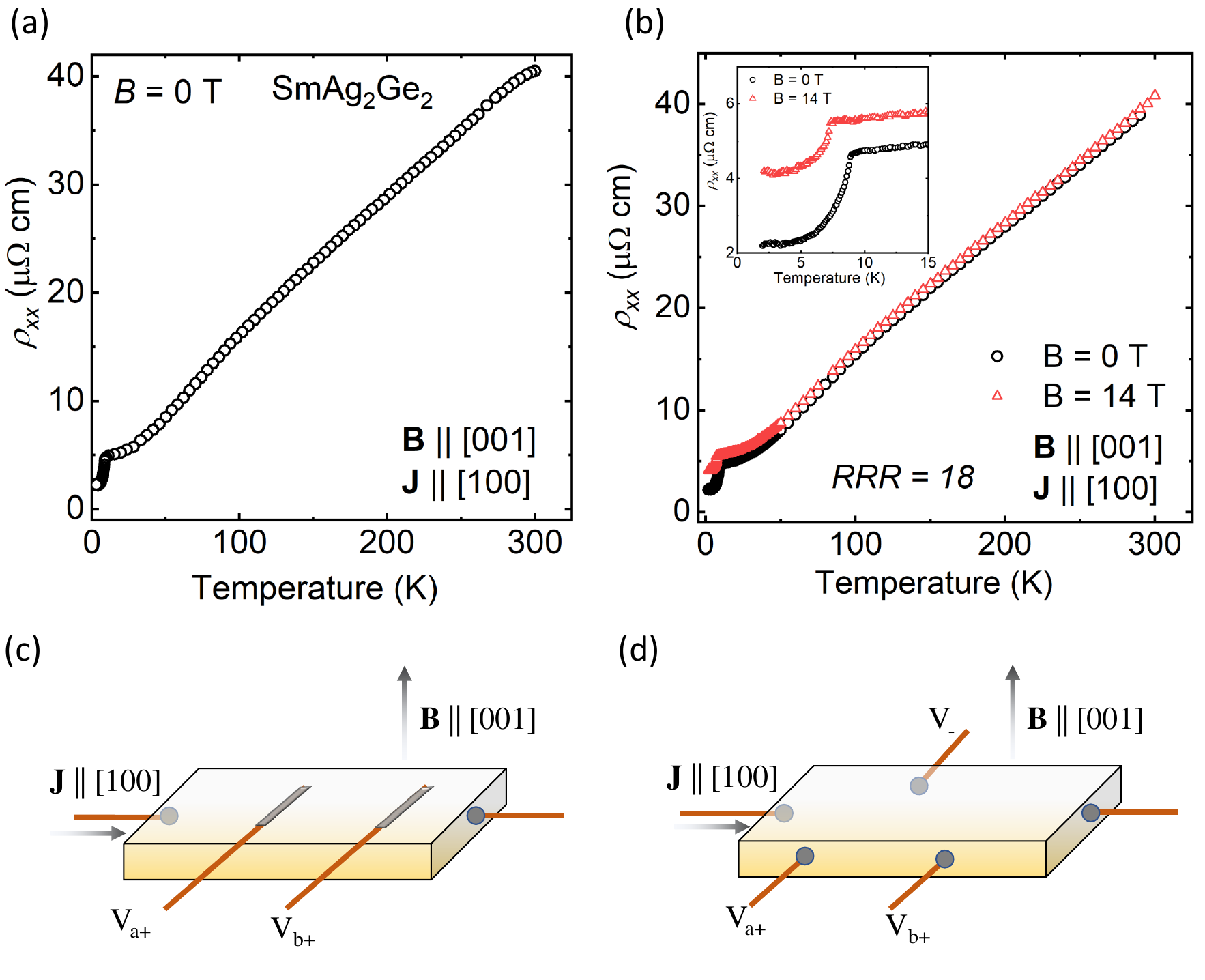}
\caption{(a) Electrical resistivity ($\rho_{xx} (T)$) as a function of the temperature for SmAg$_2$Ge$_2$. (b)  $\rho_{xx} (T)$  of SmAg$_2$Ge$_2$ measured in zero magnetic field and in a magnetic field of $14$~T. (c) Transverse and (d) Hall resistivity measurements.}
\label{fig4}
\end{figure}

The longitudinal electrical resistivity $\rho_{xx}(T)$ of SmAg$_2$Ge$_2$ as a function of temperature is depicted in Fig.~\ref{fig4}(a). The data were obtained using the standard four-probe method, as illustrated schematically in Fig.~\ref{fig4}(c). The resistivity $\rho_{xx}(T)$  decreases linearly with temperature until a sudden drop at $T_{\rm N}$, which can be attributed to a reduction in spin-disorder scattering. Below $T_{\rm N}$, $\rho_{xx}(T)$ decreases rapidly, reaching a value of $2.2~\mu \Omega~\text{cm}$ at $2$~K. The residual resistivity ratio (RRR) for the sample is estimated as $18$ and the overall behavior appeared to be metallic in nature. Notably, $\rho_{xx}(T)$ data for both cooling and heating cycles is almost identical in the paramagnetic region, demonstrating the absence of thermal hysteresis. The resistivity data exhibit typical metallic behavior for SmAg$_2$Ge$_2$.  The field-dependent $\rho_{xx} (T)$, measured in zero field and $14$~T,  are shown in Fig.~\ref{fig4}(b).  For the $14$~T field, the $T_{\rm N}$ shifts to $6.4$~K, consistent with the heat capacity measurement. The resistivity remains nearly unchanged in the paramagnetic region for both 0 and $14$~T magnetic fields.

\begin{figure}[t!]
\includegraphics[width=0.49\textwidth]{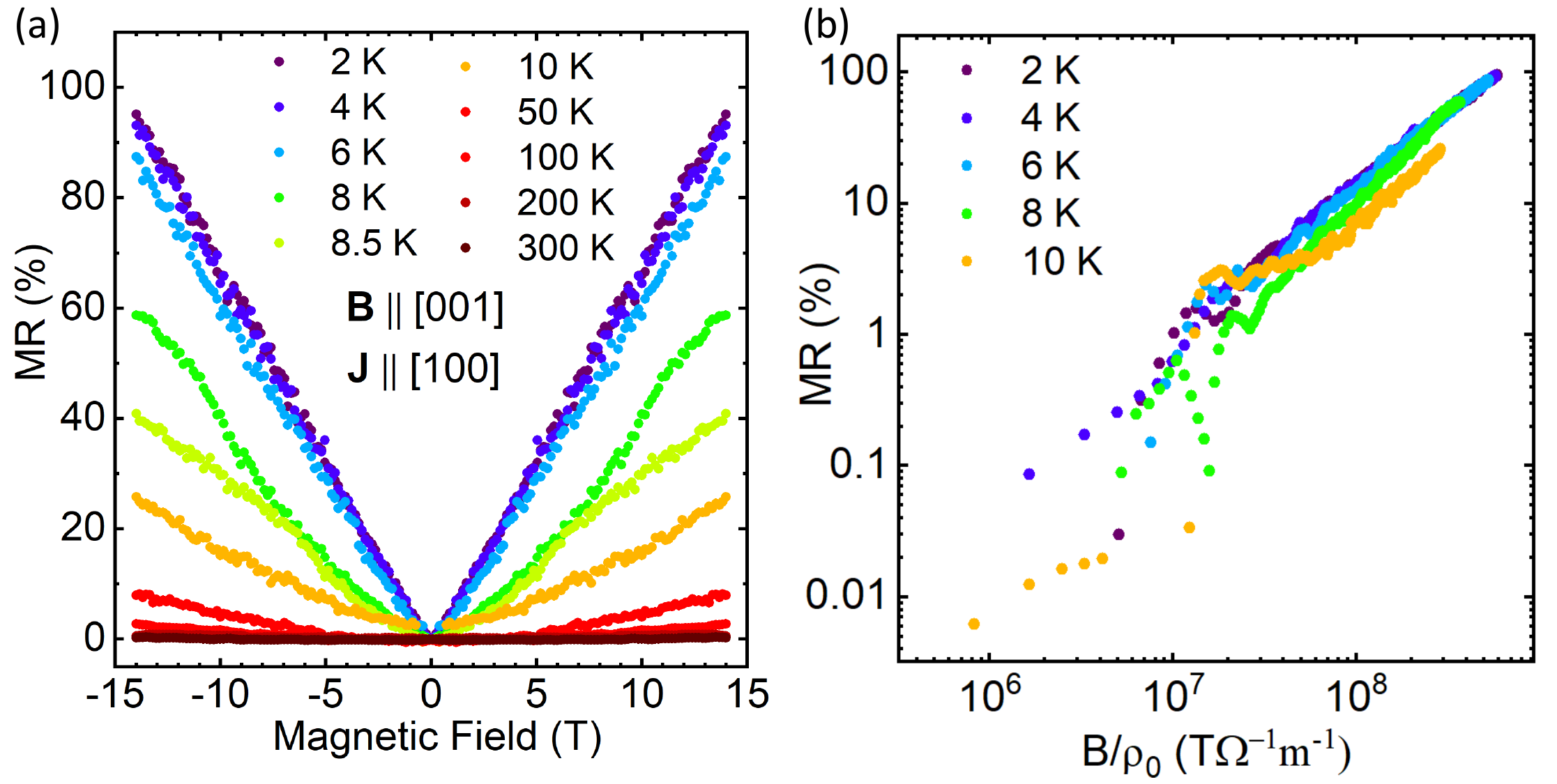}
\caption{(a) Magnetoresistance (MR) of SmAg$_2$Ge$_2$ as a function of the applied magnetic field at various temperatures. (b) Kohler’s scaling of MR at different temperatures.}
\label{fig5}
\end{figure}

The magnetoresistance (MR) at various fixed temperatures for $J$ $\parallel[100]$ and $B$ $\parallel [001]$ are shown in Fig.~\ref{fig5}(a). The MR (\%) is defined as $\left[ MR = \frac{\rho (B) - \rho (0)} {\rho (0)} \times 100 \right]$, where $\rho (B)$ and $\rho (0)$ are the resistivities at an applied magnetic field $B$ and zero field, respectively. The MR shows positive values without any sign of saturation from $2$ to $300$~K under applied magnetic fields up to $14$~T.  Importantly, the MR reaches a value of $\sim 97$\%  at low temperatures, indicating a relatively strong response of the sample to the applied magnetic field. The observed positive MR at all temperatures indicates an enhancement of spin scattering due to the magnetic field. Moreover, the MR exhibits temperature-dependent behavior. It increases linearly with the magnetic field at low temperatures of $2$, $4$, and $6$~K, displaying a nearly symmetrical $V$-shaped curve, suggesting a linear magnetoresistance (LMR) behaviour. Above $8$~K, the MR exhibits a slight parabolic bend in the low-field limit.

At high temperatures $T~>~50$~K, the MR follows a quadratic behaviour which is mainly attributed to the Lorentz force acting on the charge carriers. The LMR at low $T$ may arise due to various factors considering the complex electronic structure and multi-pocket Fermi surface of SmAg$_2$Ge$_2$ (see band structure details below). Typically, MR follows a $B^{2}$ dependence for compensated semimetals; however, this is not the case for SmAg$_2$Ge$_2$. To investigate the underlying behavior of MR, we plot MR versus $B/\rho_0$ on a logarithmic scale for different temperatures in Fig.~\ref{fig5}(b). It is obvious from the figure that the MR does not fall on to a single curve, signifying the fact that the Kohler's rule breaks down due to different scattering mechanisms at different temperatures.

The relatively low effective mass ($0.47~m_0$ obtained from quantum oscillations studies~-~\textit{vide infra}) results in a high cyclotron frequency which leads to a dominant linear contribution to the magnetoresistance, overshadowing the quadratic magnetoresistance arising from the various parts of the Fermi surface. We emphasize that, given the presence of linear band crossings in SmAg$_2$Ge$_2$ near the Fermi level, Abrikosov's quantum description of LMR could be applicable. According to the Abrikosov's model ~\cite{AAAbrikosov_2000}, when the charge carriers are confined in the lowest Landau level and reach the quantum limit the LMR can be expressed as, 

\begin{equation}\label{eq2}
\rho_{xx} = \frac{N_i B}{\pi c e N_2^2}
\end{equation}

where $N_i$ is the static scattering center density and $n_e$ is the carrier concentration. The applicability of this quantum LMR is valid if $N_i~\ll~n_e$ and the temperature $T$ should satisfy the following expression so that it is smaller than the Landau level spacing:
\begin{equation}\label{eq3}
T_{\rm limit} < \frac{e B \hbar}{m^* k_{\rm B}}
\end{equation}
where the physical constants have their usual meaning and are considered in SI units. With the effective mass as $0.47~m_0$ and for a magnetic field of $14$~T, $T_{\rm limit}$ is estimated as $40$~K.  Our experimental results agree well with this $T$ limit, where the LMR is observed for $T~<~40$~K, and above this temperature a non-linear/quadratic behavior is observed. The experimental data, however, show that the LMR decreases with increasing temperature. Another widely recognized mechanism for explaining LMR is the Parish-Littlewood (PL) model~\cite{Parish2003, PhysRevB.58.2788, PhysRevB.75.214203}, which suggests that linear MR is influenced by carrier mobility and disorder effects. Since SmAg$_2$Ge$_2$ is not a disordered system, the PL model is not applicable here. The coexistence of a complex band structure, including multiple linear band crossings at the Fermi level and hybridized flat bands, may account for the complex behavior of MR and warrants further investigations~\cite{feng2015large, zhou2020linear}.

\subsection{Anomalous Hall effect in SmAg$_2$Ge$_2$}

\begin{figure*}[ht!]
\includegraphics[width=0.85\textwidth]{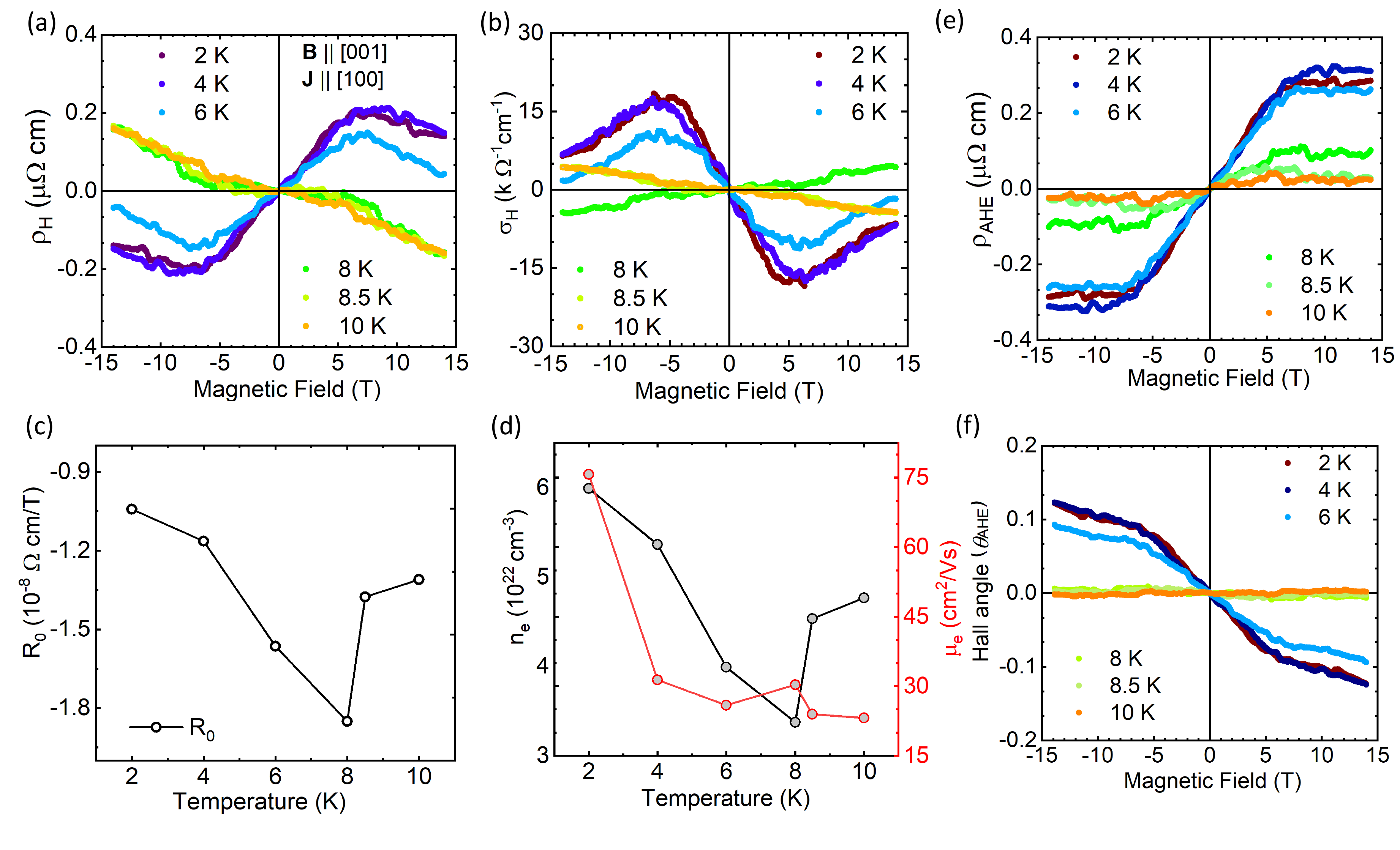}
\caption{(a) Hall resistivity and (b) Hall conductivity of SmAg$_2$Ge$_2$ as a function of the magnetic field ($B$) up to $14$~T at various temperatures. (c) Normal Hall coefficient  $R_0$ and (d) carrier concentration (left axis) and Hall mobility (right axis) as a function of temperature. (e) Anomalous Hall resistivity and (f) anomalous  Hall angle of SmAg$_2$Ge$_2$ as a function of magnetic field at various temperatures.}
\label{fig6}
\end{figure*}

 To estimate the carrier concentration, we present the Hall resistivity  $\rho_H$ of SmAg$_2$Ge$_2$ as a function of magnetic field for various temperatures in Fig.~\ref{fig6}(a). The Hall measurements were performed using a standard five-probe geometry (see Fig.~\ref{fig4}(d)), with the magnetic field applied along the crystallographic $c$-axis. The Hall data were antisymmetrized to eliminate contributions from linear resistivity $\rho_{xx}$. The most prominent feature of  $\rho_H$ versus $B$ data is its deviation from a linear behavior, exhibiting a broad peak at temperatures 2, 4, and $6$~K below $B\sim 10$~T.  The peak magnetic field is $\sim 8$~T at $2$ and $4$~K, while it shifts slightly to a lower value at $6$~K. Beyond the peak field,  $\rho_H$ gradually transitions to a linear behavior. At $8$~K and higher temperatures, no peak-like feature is observed, instead, a linear relationship between $\rho_H$ and the magnetic field is established. 

 In Fig.~\ref{fig6}(b), we present the calculated Hall conductivity $\sigma_{H}=\frac{\rho_{H}}{\rho_{xx}^2+\rho_{H}^2}$, where $\rho_{xx}$ and $\rho_{H}$ are the linear and Hall resistivities, respectively. The electron concentration and mobility are inferred from the measurements of $\rho_H$ and $\sigma_H$ in Fig.~\ref{fig6}(d) by assuming a single band picture at different temperatures. The analysis of $\rho_H$ and $\sigma_H$ indicates that a single band model adequately describes the carriers in SmAg$_2$Ge$_2$ (details are given in Supplementary Material~\cite{SM}). Considering the low-temperature behavior of $\rho_H$ below $8$~T, we also tried two-band model fitting using both $\rho_H$ and $\rho_{xx}$. However, simultaneous fitting to the two-band model is not feasible.
 
 The carrier density $n_e$ and Hall mobility $\mu_e$ are derived from the $\rho_H$ using the relations $n_e = \frac{1}{eR_0}$ and $ \mu_e = \frac{R_0}{\rho (B = 0)}$, where $R_0$ is the slope of the $\rho_{H}$ curves at higher magnetic field. The calculated values of $R_0$ as a function of temperature are shown in Fig.~\ref{fig6}(c). The negative values of $R_0$ at all temperatures indicate that electrons are majority carriers. The calculated values of $n_e$ and $\mu_e$ are presented in Fig.~\ref{fig6}(d). The carrier concentration $n_e$ remains 10$^{22}$~cm$^{-3}$ with minimal temperature dependence.
 
To investigate the presence of the anomalous Hall effect in SmAg$_2$Ge$_2$ under an applied magnetic field, we present anomalous Hall resistivity $\rho_{\text{AHE}}$ in Fig.~\ref{fig6}(e). In metallic systems exhibiting spontaneous magnetization, the total Hall resistivity can be expressed as $\rho_{H} = \rho^{\text{N}}_{H} + \rho^{\text{AHE}}_{H} = R_0B + R_sM$, where $R_0$ is the normal Hall coefficient, $R_s$ is anomalous Hall coefficient, and $M$ is the magnetization. The anomalous Hall resistivity $\rho^{\text{AHE}}$ increases rapidly with the magnetic field up to $8$~T, after which it changes more slowly for 2, 4, and $6$~K temperatures. In contrast, $\rho^{\text{AHE}}$ shows weak dependence on the magnetic field at higher temperatures. We further calculate the Hall angle for SmAg$_2$Ge$_2$. The anomalous Hall angle is defined as the, $\theta_\text{AHE}=\text{tan}^{-1}(\frac{\sigma_H^{AHE}}{\sigma_{xx}})$, where $\sigma_H^{AHE}$ and $\sigma_{xx}$ are anomalous Hall conductivity and longitudinal conductivity, respectively. As illustrated in Fig.~\ref{fig6}(f), the anomalous Hall angle exhibits an unusual dependence on the magnetic field. The Hall angle reaches a value of 0.14 at $10$~T for $2$~K and decreases with increasing temperature. This value is either comparable to or exceeds those previously reported in antiferromagnetic systems~\cite{shekhar2018anomalous,zhu2020exceptionally,suzuki2016large}. 

The anomalous Hall effect in SmAg$_2$Ge$_2$ appears within the antiferromagnetic state under an applied magnetic field below the $T_{\rm N}$ = 9.2~K. The AHE  may originate due to various effects such as: intrinsic Berry curvature effects, noncollinear spin textures, and extrinsic mechanisms such as skew scattering and side jump effects. Because intrinsic Berry curvature is independent of scattering processes, we expect the anomalous Hall effect to be temperature-independent within the antiferromagnetic region, as seen in our results. However, in the antiferromagnetic state under a magnetic field, spin canting occurs, potentially introducing noncollinear spin structures that could influence the anomalous Hall effect. We further investigate the intrinsic Berry curvature effects in the band structure section below.

\subsection{de Haas-van Alphen oscillations}

\begin{figure}[!]
\includegraphics[width=0.49\textwidth]{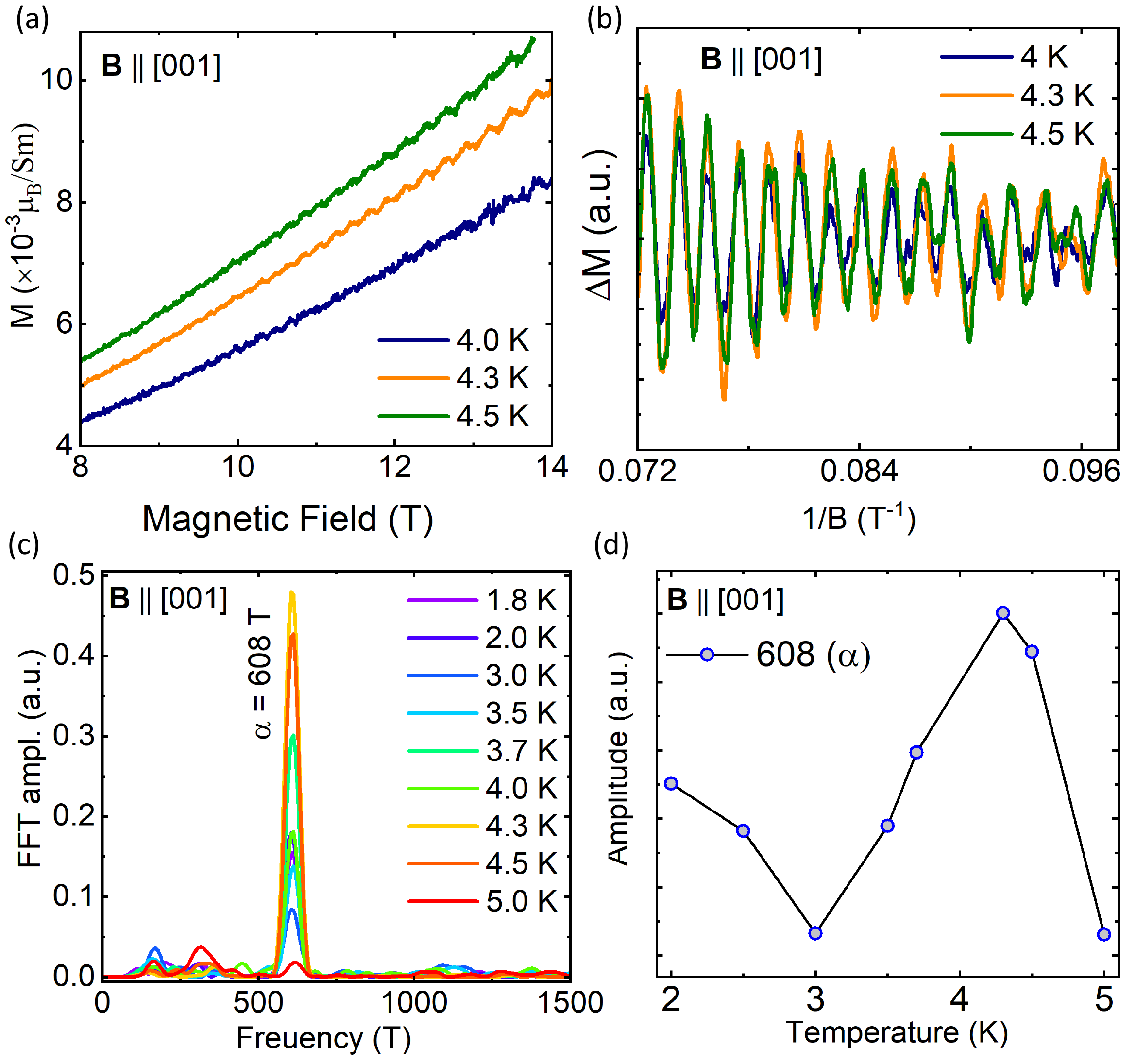}
\caption{(a) Magnetization $M$ and (b) background-substracted $\Delta M$ of SmAg$_2$Ge$_2$ as a function of the magnetic field B at various temperatures for $B$ $\parallel [001]$. The de-Haas-van Alpen (dHvA) oscillations are observed for $B~>~8$~T. (c) Fast-Fourier-Transform (FFT) frequency spectrum of  SmAg$_2$Ge$_2$ at different temperatures. (d) Temperature dependence of FFT amplitude of the $\alpha$ = $608$~T oscillation.}
\label{fig7}
\end{figure}

We now discuss the de Haas-van Alphen effect (dHvA) effect oscillations in the magnetization of SmAg$_2$Ge$_2$. Figure~\ref{fig7}(a) presents the magnetization as a function of $B$ along the [001] direction. The magnetization exhibits pronounced dHvA oscillations for $B~>~10$~T at $4.3$~K. As the temperature increases, these oscillations diminish and become indistinguishable at temperatures above $5$~K. To isolate the oscillatory component of magnetization, we subtracted the non-oscillatory background data (see Fig.~\ref{fig7}(b)) and extracted the dHvA frequency by fast Fourier transform (FFT). The FFT spectrum as a function of temperature is shown in Fig.~\ref{fig7}(c). It reveals a single fundamental frequency $\alpha$ = $608$~T, which initially decreases with temperature until $3$~K and then increases, peaking at $4.3$~K. With a further increase in temperature, the amplitude of the frequency decreases. This behavior is more clear in the temperature dependence of the FFT amplitude in Fig.~\ref{fig7}(d).  Due to this nonmonotonic temperature behavior,  dHvA data do not follow the Lifshitz-Kosevich (LK) formula. Such deviations might arise from challenges in disentangling the magnetization contributions from different Fermi surface sheets.

\begin{figure}[t!]
\includegraphics[width=0.49\textwidth]{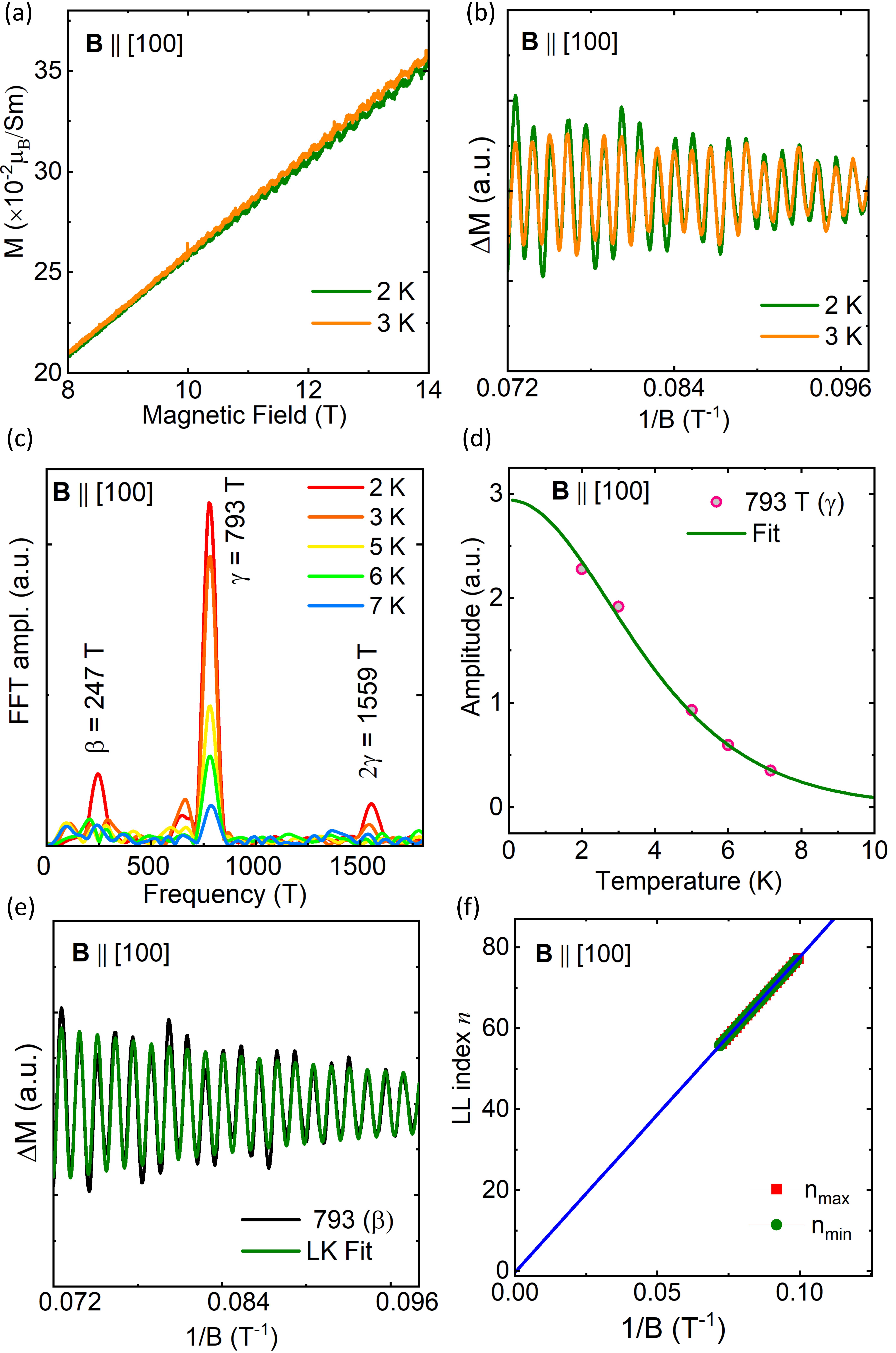}
\caption{(a) Magnetization $M$ and (b) background-substracted $\Delta M$ of SmAg$_2$Ge$_2$ as a function of the magnetic field $\textit{\textbf{B}}$ at $T$~=~$2$ and $3$~K for $B$ $\parallel [100]$. de Haas-van Alphen oscillations are observed for $B~>~10$~T. (c) The FFT frequency spectrum of  SmAg$_2$Ge$_2$ at different temperatures. (d) Temperature dependence of FFT amplitude of the $\gamma$ = $793$~T oscillation. The solid line represents the fit to the thermal damping factor of Lifshitz-Kosevich (LK) expression. (e) Band-pass filtered dHvA oscillation with frequency F$\alpha$ = $793$~T (black line) and LK fitting (green line) at $2$~K. (f) Landau level (LL) fan diagram corresponding to $\gamma$ = $793$~T.}
\label{fig8}
\end{figure}

In Fig.~\ref{fig8}, we present dHvA oscillations for SmAg$_2$Ge$_2$ when magnetic field is applied along $B~\parallel$~[100]. Figure~\ref{fig8}(a) shows magnetization as a function of the field at various temperatures. Similar to the case of $B$ $\parallel [001]$, the magnetization exhibits oscillations for fields above $10$~T at lower temperatures. The dHvA oscillations are more clearly resolved in background substrated magnetization data shown in Fig.~\ref{fig8}(b) and the FFT spectrum presented in  Fig.~\ref{fig8}(c). Along this direction, we extracted two fundamental frequencies $\beta$ = $247$~T and $\gamma$ = $793$~T with a second harmonic at $2\gamma$= $1559$~T. Notably, $\alpha$ = $608$~T frequency observed for $B$ $\parallel [001]$ is not seen along this direction. The temperature dependence of FFT amplitude (Fig.~\ref{fig8}(c)) shows that the amplitude of $\alpha$ frequency is visible only at $2$~K and diminishes for other temperatures, whereas the amplitude of $\gamma$ frequency decreases monotonically with increasing temperature. The temperature dependence of the $\gamma$ frequency amplitude is fitted to the thermal damping factor $R_T$ of the LK formula, 
\begin{equation}
\Delta M \propto -R_T R_D B^{k} \sin{\left[ 2\pi \left( \frac{F}{B} + \psi \right) \right]}
\label{LKf} 
\end{equation}   
where $R_T= (\lambda T \mu/B)/(\text{sinh}(\lambda T \mu/B)$, $R_D=\text{exp}(-\lambda T_D \mu/B)$, and $\lambda = (2 \pi^2 k_B m_0)/(\hbar e)$. Here $\mu$ is the ratio of effective mass $m^*$ to free electron mass $m_0$, and $T_D$ is the Dingle temperature.  In Eq.~\ref{LKf}, $k$ is $1/2$ for a three-dimensional (3D) Fermi surface and zero for a 2D Fermi surface. The phase factor $\psi$ is given by $\psi = \left[ (\frac{1}{2}-\frac{\Phi_B}{2\pi})-\delta \right]$, where $\Phi_B$ is the Berry phase and $\delta$ is an additional phase factor which depends on the dimensionality of the Fermi surface. The fitting of amplitude data to $R_T$ (Fig.~\ref{fig8}(d)) yields an effective mass of $m^*$=0.47~$m_0$ for the $\gamma$ frequency. 
 
To further characterize the $\gamma$ pocket, we estimate various parameters in the LK formula in Eq.~\ref{LKf}. Applying the Onsagar relation $\frac{\hbar A_i}{2\pi e}$, we find the cross-section of  the $\gamma$ pocket to be $7.558$ nm${^{-2}}$, with a Fermi wave vector $k_{\rm F}$ of $1.540\times10^7 \text{cm}^{-1}$. The corresponding Fermi velocity $v_{\rm F}$ is obtained as $3.453\times10^7 \text{cm~s}^{-1}$. To estimate the Dingle temperature $T_{\rm D}$ for the $\gamma$ pocket, we used a bandpass filter to isolate the oscillation corresponding to the frequency 793~T (Fig.~\ref{fig8}(e)). The estimated $T_{\rm D}$ is 1.25~K, considering this pocket as a 3D Fermi pocket. The scattering lifetime $\tau$ is determined to be $9.72 \times 10^{-13}$s and quantum mobility is 3063~cm$^2$V$^{-1}$s$^{-1}$. We also extract the Berry phase associated with the Fermi pocket by plotting the Landau level (LL) fan diagram. We assign the LL-index $n\pm\frac{1}{4}$~($=F/B+\psi$) to the maxima and minima of quantum oscillations for $+$ and $-$ sign, respectively. Employing the bandpass-filtered oscillation data, we construct the LL fan diagram and plot $n$ as a function of the inverse magnetic field in Fig.~\ref{fig8}(f). The intercept is -0.328, leading to a calculated Berry phase of $-0.41\pi$, which indicates the trivial nature of the Fermi pocket.  
  
\subsection{Electronic structure and anomalous Hall conductivity}

We now discuss the electronic properties and calculated anomalous Hall conductivity of SmAg$_2$Ge$_2$. SmAg$_2$Ge$_2$ exhibits an antiferromagnetic state below $T_{\rm N}$ = 9.2~K, and hence we calculate the band structure in both paramagnetic and antiferromagnetic states. The band structure of the paramagnetic state is modeled using a nonmagnetic unit cell, treating Sm-$f$ electrons as core electrons. The antiferromagnetic bands are calculated in the conventional tetragonal unit cell containing two Sm atoms per unit cell (as shown in Fig.~\ref{fig1}(a)) and unfolded into a $1\times 1\times1$ primitive Brillouin zone for direct comparison.  
 
\begin{figure}[t!]
\includegraphics[width=0.49\textwidth]{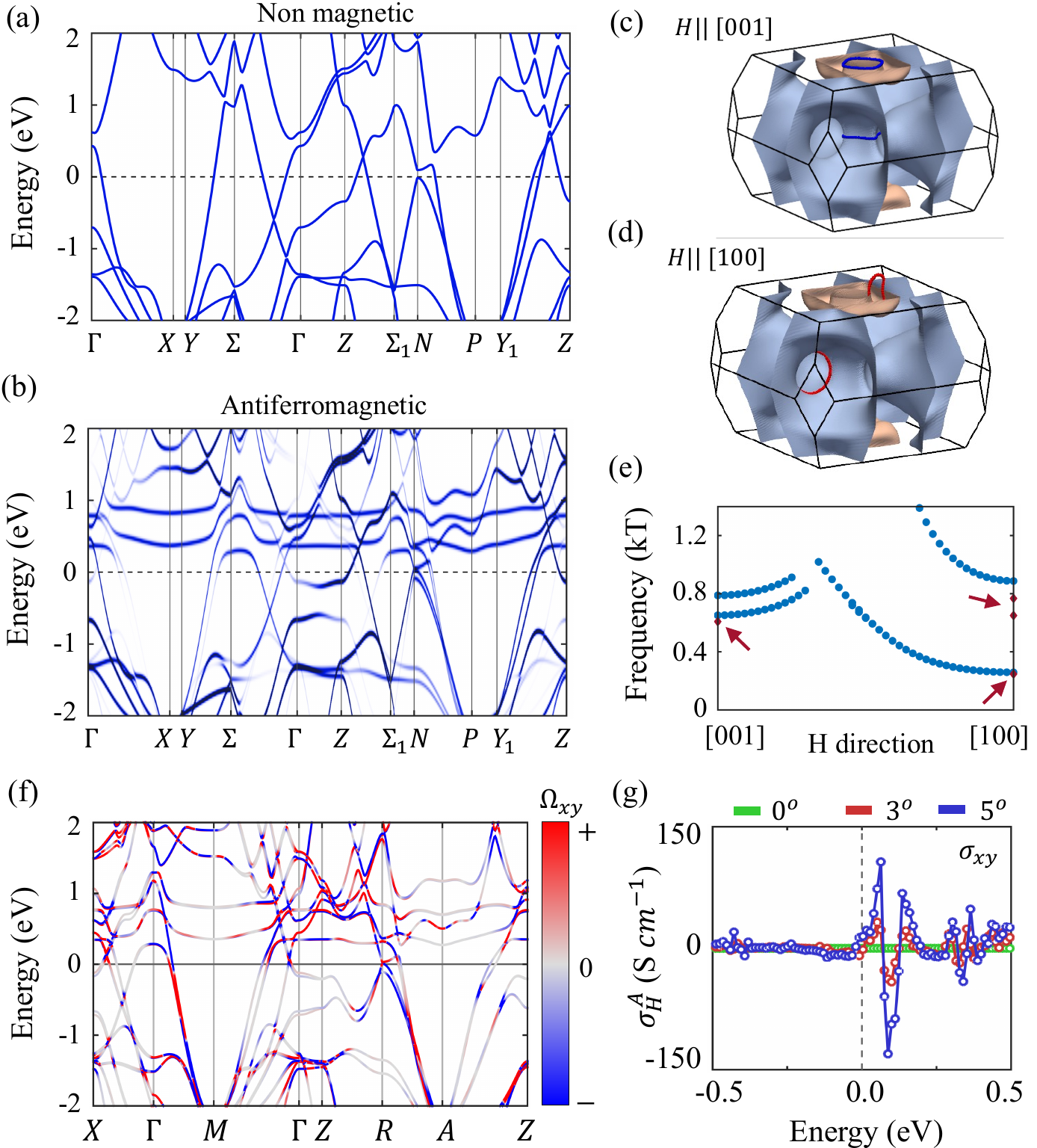}
\caption{Bulk band structure of SmAg$_2$Ge$_2$ in (a) nonmagnetic and (b) antiferromagnetic state along the high-symmetry direction in the primitive cell Brillouin zone. The antiferromagnetic state band structure is obtained in the conventional unit cell and subsequently unfolded into the primitive Brillouin zone. (c)-(d) Calculated Fermi surface of SmAg$_2$Ge$_2$. Blue and red contours represent carrier orbits when the magnetic field is applied along [001] and [100] directions, respectively. (e) Calculated angle-dependent oscillation frequencies associated with the blue and red contours shown in (c) and (d). Experimental data points are indicated by filled red circles. (f) Berry curvature resolved band structure for 5~$^{\circ}$ spin-canting angle with respect to the antiferromagnetic state. The color bar represents the Berry curvature. (g) The $\sigma_{xy}$ component of intrinsic anomalous Hall conductivity of SmAg$_2$Ge$_2$ as a function of energy for a spin-canting angle 0~$^{\circ}$, 3~$^{\circ}$, and 5~$^{\circ}$. The vertical dashed line marks the Fermi level.}
\label{fig_t1}
\end{figure}

Figure~\ref{fig_t1}(a) presents the calculated band structure of SmAg$_2$Ge$_2$ in the nonmagnetic state, revealing a metallic character where various bands with nearly linear energy dispersion cross the Fermi level. In the presence of antiferromagnetic ordering (Fig.~\ref{fig_t1}(b)), the band structure retains a similar metallic energy dispersion. Importantly, the number of bands crossing the Fermi level remains the same in both nonmagnetic and antiferromagnetic states, with nearly identical band dispersion near the Fermi level. The Sm$-f$ states appear above the Fermi level, around 1~$eV$. While antiferromagnetic ordering introduces band folding and hybridization due to the emergence of new potential, these effects are more pronounced away from the Fermi level, without changing the overall Fermi surface of the nonmagnetic state. Figures~\ref{fig_t1}(c) and \ref{fig_t1}(d) illustrate the calculated Fermi surface of SmAg$_2$Ge$_2$, which consists of multiple Fermi pockets. We calculate the quantum oscillations associated with these Fermi pockets and find a reasonable match with the observed oscillation frequencies in experiments. Notably, the blue and red contours in Figs.~\ref{fig_t1}(c) and \ref{fig_t1}(d) represent the electron orbits for the applied magnetic field along the [001] and [100] directions, respectively, which fall within the experimentally observed frequency range. Specifically, when the field is along [001], the oscillation frequencies are $651$~T and 788~T, whereas for the [100] direction, the frequencies are $259$~T and $887$~T. The calculated oscillation frequencies for the [100] and [001] field directions are presented in Fig.~\ref{fig_t1}(f), with experimentally observed frequencies indicated by red dots and arrows. These results demonstrate good agreement with our experimental findings.

As discussed in the Hall results, SmAg$_2$Ge$_2$ exhibits an anomalous Hall effect under an applied magnetic field in the antiferromagnetic region. Assuming that the magnetic field modifies its antiferromagnetic spin structure and introduces spin canting, we calculate the band structure, Berry curvature, and anomalous Hall conductivity in the canted antiferromagnetic state. The spin canting angle is defined as the deviation of the spin-down moment from the negative $z-$axis in the presence of $B$ $\parallel [001]$. Figure~\ref{fig_t1}(f) presents the calculated Berry-curvature resolved band structure of SmAg$_2$Ge$_2$ with a spin-canting angle of 5~$^{\circ}$. The spin canting clearly results in a non-zero Berry curvature in the band structure, as indicated in the color scale. Importantly, due to the presence of inversion $\mathcal{I}$ and effective time-reversal symmetry $\mathcal{S} = \Theta \tau_{1/2}$ (where $\Theta$ is the time-reversal operator and $\tau_1/2$ is the half-translation vector connecting spin-up and spin-down Sm atoms), the Berry curvature is not allowed in perfectly antiferromagnetic state. The spin canting breaks effective time-reversal symmetry, thereby enabling a non-zero Berry curvature. Figure~\ref{fig_t1}(g) presents the calculated anomalous Hall conductivity from Berry curvature using the Kubo formula~\cite{AHC}. The in-plane anomalous Hall conductivity $\sigma_{xy}$ as a function of energy is shown for three different canting angles (0$^{\circ}$, 3$^{\circ}$, and 5$^{\circ}$). As seen in experimental results, the anomalous Hall conductivity is zero in the antiferromagnetic state (no canting). However, as the canting angle increases to 3$^{\circ}$ and 5$^{\circ}$, the magnitude of $\sigma_{xy}$ starts increasing at the Fermi level, peaking at $E_f+ 60$~meV. This peak grows from 0 to 130~S cm$^{-1}$ as the canting angles change from 0$^{\circ}$ to 5$^{\circ}$. These results indicate that SmAg$_2$Ge$_2$ can exhibit a Berry curvature-driven anomalous Hall effect. However, the magnitude found in our calculations is lower than the experimental results, indicating that other contributing mechanisms are likely involved.
\\
\section{Conclusion}

We present a comprehensive analysis of magnetotransport and Fermi surface properties of as-grown single crystals of the rare-earth intermetallic SmAg$_2$Ge$_2$. Our results indicate that SmAg$_2$Ge$_2$ crystallizes in a ThCr$_2$Si$_2$-type tetragonal structure and displays antiferromagnetic order below $T_{\rm N}$ = 9.2~K, with the easy axis oriented along the tetragonal $c$-axis. The antiferromagnetic transition is clearly delineated in specific heat and the entropy calculation reveals a doublet ground state. Electrical measurements confirm that SmAg$_2$Ge$_2$ is metallic. Field-dependent magnetotransport measurements demonstrate a significant non-saturating magnetoresistance, reaching approximately $\sim 97$\% at 2~K when the magnetic field is applied parallel to the $c$-axis. Notably, the magnetoresistance exhibits linear behavior in the antiferromagnetic state, transitioning to a quadratic dependence at higher temperatures, for $T~>~40$~K. Hall effect measurements indicate that electrons are the majority carriers with high mobility, along with a significant anomalous Hall effect observed at 2, 4, and $6$~K. We further analyze dHvA oscillations for magnetic fields applied in both in-plane and out-of-plane directions and discuss the Fermi surface and associated nontrivial behaviors. Our first-principles results of nonmagnetic, antiferromagnetic, and canted-antiferromagnetic states are in reasonable agreement with the experimental results. This study establishes SmAg$_2$Ge$_2$ as an interesting rare-earth intermetallic material, which can offer a unique platform for investigating nontrivial magnetotransport properties influenced by its magnetic ordering and complex electronic structure.

\section*{Acknowledgements}
K. B. acknowledges the Department of Science and Technology, Government of India for financial support through WISE-PDF research project No. DST/WISE-PDF/PM-2/2023(G). This work is supported by the Department of Atomic Energy of the Government of India under Project No. 12-R$\&$D-TFR-5.10-0100 and benefited from the HPC resources of TIFR Mumbai. 


\begin{thebibliography}{47}%
	\makeatletter
	\providecommand \@ifxundefined [1]{%
		\@ifx{#1\undefined}
	}%
	\providecommand \@ifnum [1]{%
		\ifnum #1\expandafter \@firstoftwo
		\else \expandafter \@secondoftwo
		\fi
	}%
	\providecommand \@ifx [1]{%
		\ifx #1\expandafter \@firstoftwo
		\else \expandafter \@secondoftwo
		\fi
	}%
	\providecommand \natexlab [1]{#1}%
	\providecommand \enquote  [1]{``#1''}%
	\providecommand \bibnamefont  [1]{#1}%
	\providecommand \bibfnamefont [1]{#1}%
	\providecommand \citenamefont [1]{#1}%
	\providecommand \href@noop [0]{\@secondoftwo}%
	\providecommand \href [0]{\begingroup \@sanitize@url \@href}%
	\providecommand \@href[1]{\@@startlink{#1}\@@href}%
	\providecommand \@@href[1]{\endgroup#1\@@endlink}%
	\providecommand \@sanitize@url [0]{\catcode `\\12\catcode `\$12\catcode
		`\&12\catcode `\#12\catcode `\^12\catcode `\_12\catcode `\%12\relax}%
	\providecommand \@@startlink[1]{}%
	\providecommand \@@endlink[0]{}%
	\providecommand \url  [0]{\begingroup\@sanitize@url \@url }%
	\providecommand \@url [1]{\endgroup\@href {#1}{\urlprefix }}%
	\providecommand \urlprefix  [0]{URL }%
	\providecommand \Eprint [0]{\href }%
	\providecommand \doibase [0]{https://doi.org/}%
	\providecommand \selectlanguage [0]{\@gobble}%
	\providecommand \bibinfo  [0]{\@secondoftwo}%
	\providecommand \bibfield  [0]{\@secondoftwo}%
	\providecommand \translation [1]{[#1]}%
	\providecommand \BibitemOpen [0]{}%
	\providecommand \bibitemStop [0]{}%
	\providecommand \bibitemNoStop [0]{.\EOS\space}%
	\providecommand \EOS [0]{\spacefactor3000\relax}%
	\providecommand \BibitemShut  [1]{\csname bibitem#1\endcsname}%
	\let\auto@bib@innerbib\@empty
	\bibitem [{\citenamefont {Buschow}(1977)}]{Buschow_1977}%
	\BibitemOpen
	\bibfield  {author} {\bibinfo {author} {\bibfnamefont {K.~H.~J.}\
			\bibnamefont {Buschow}},\ }\bibfield  {title} {\bibinfo {title}
		{{Intermetallic compounds of rare-earth and 3$d$ transition metals}},\ }\href
	{https://doi.org/10.1088/0034-4885/40/10/002} {\bibfield  {journal} {\bibinfo
			{journal} {Rep. Prog. Phys.}\ }\textbf {\bibinfo {volume} {40}},\ \bibinfo
		{pages} {1179} (\bibinfo {year} {1977})}\BibitemShut {NoStop}%
	\bibitem [{\citenamefont {Pfleiderer}(2009)}]{RevModPhys.81.1551}%
	\BibitemOpen
	\bibfield  {author} {\bibinfo {author} {\bibfnamefont {C.}~\bibnamefont
			{Pfleiderer}},\ }\bibfield  {title} {\bibinfo {title} {Superconducting phases
			of $f$-electron compounds},\ }\href
	{https://doi.org/10.1103/RevModPhys.81.1551} {\bibfield  {journal} {\bibinfo
			{journal} {Rev. Mod. Phys.}\ }\textbf {\bibinfo {volume} {81}},\ \bibinfo
		{pages} {1551} (\bibinfo {year} {2009})}\BibitemShut {NoStop}%
	\bibitem [{\citenamefont {Lai}\ \emph {et~al.}(2022)\citenamefont {Lai},
		\citenamefont {Chan},\ and\ \citenamefont {Baumbach}}]{LaiSciAv2022}%
	\BibitemOpen
	\bibfield  {author} {\bibinfo {author} {\bibfnamefont {Y.}~\bibnamefont
			{Lai}}, \bibinfo {author} {\bibfnamefont {J.~Y.}\ \bibnamefont {Chan}},\ and\
		\bibinfo {author} {\bibfnamefont {R.~E.}\ \bibnamefont {Baumbach}},\
	}\bibfield  {title} {\bibinfo {title} {{Electronic landscape of the
				$f$-electron intermetallics with the ThCr$_2$Si$_2$ structure}},\ }\href
	{https://doi.org/10.1126/sciadv.abp8264} {\bibfield  {journal} {\bibinfo
			{journal} {Sci. Adv.}\ }\textbf {\bibinfo {volume} {8}},\ \bibinfo {pages}
		{eabp8264} (\bibinfo {year} {2022})}\BibitemShut {NoStop}%
	\bibitem [{\citenamefont {Ogunbunmi}\ \emph {et~al.}(2021)\citenamefont
		{Ogunbunmi}, \citenamefont {Baranets},\ and\ \citenamefont
		{Bobev}}]{Ogunbunmi2021}%
	\BibitemOpen
	\bibfield  {author} {\bibinfo {author} {\bibfnamefont {M.~O.}\ \bibnamefont
			{Ogunbunmi}}, \bibinfo {author} {\bibfnamefont {S.}~\bibnamefont
			{Baranets}},\ and\ \bibinfo {author} {\bibfnamefont {S.}~\bibnamefont
			{Bobev}},\ }\bibfield  {title} {\bibinfo {title} {{Synthesis and Transport
				Properties of the Family of Zintl Phases Ca$_3$RESb$_3$ (RE = La--Nd, Sm,
				Gd--Tm, Lu): Exploring the Roles of Crystallographic Disorder and Core 4f
				Electrons for Enhancing Thermoelectric Performance}},\ }\href
	{https://doi.org/10.1021/acs.chemmater.1c03300} {\bibfield  {journal}
		{\bibinfo  {journal} {Chem. Mater.}\ }\textbf {\bibinfo {volume} {33}},\
		\bibinfo {pages} {9382} (\bibinfo {year} {2021})}\BibitemShut {NoStop}%
	\bibitem [{\citenamefont {Haule}\ and\ \citenamefont
		{Kotliar}(2009)}]{HiddenOrder_URu2Si2}%
	\BibitemOpen
	\bibfield  {author} {\bibinfo {author} {\bibfnamefont {K.}~\bibnamefont
			{Haule}}\ and\ \bibinfo {author} {\bibfnamefont {G.}~\bibnamefont
			{Kotliar}},\ }\bibfield  {title} {\bibinfo {title} {Arrested {Kondo} effect
			and hidden order in {URu$_2$Si$_2$}},\ }\href
	{https://doi.org/10.1038/nphys1392} {\bibfield  {journal} {\bibinfo
			{journal} {Nat. Phys.}\ }\textbf {\bibinfo {volume} {5}},\ \bibinfo {pages}
		{796–799} (\bibinfo {year} {2009})}\BibitemShut {NoStop}%
	\bibitem [{\citenamefont {Villaume}\ \emph {et~al.}(2008)\citenamefont
		{Villaume}, \citenamefont {Bourdarot}, \citenamefont {Hassinger},
		\citenamefont {Raymond}, \citenamefont {Taufour}, \citenamefont {Aoki},\ and\
		\citenamefont {Flouquet}}]{Superconductivity_URu2Si2}%
	\BibitemOpen
	\bibfield  {author} {\bibinfo {author} {\bibfnamefont {A.}~\bibnamefont
			{Villaume}}, \bibinfo {author} {\bibfnamefont {F.}~\bibnamefont {Bourdarot}},
		\bibinfo {author} {\bibfnamefont {E.}~\bibnamefont {Hassinger}}, \bibinfo
		{author} {\bibfnamefont {S.}~\bibnamefont {Raymond}}, \bibinfo {author}
		{\bibfnamefont {V.}~\bibnamefont {Taufour}}, \bibinfo {author} {\bibfnamefont
			{D.}~\bibnamefont {Aoki}},\ and\ \bibinfo {author} {\bibfnamefont
			{J.}~\bibnamefont {Flouquet}},\ }\bibfield  {title} {\bibinfo {title}
		{Signature of hidden order in heavy fermion superconductor {URu$_2$Si$_2$}:
			Resonance at the wave vector {Q}$_{0}$=(1,0,0)},\ }\href
	{https://doi.org/10.1103/PhysRevB.78.012504} {\bibfield  {journal} {\bibinfo
			{journal} {Phys. Rev. B}\ }\textbf {\bibinfo {volume} {78}},\ \bibinfo
		{pages} {012504} (\bibinfo {year} {2008})}\BibitemShut {NoStop}%
	\bibitem [{\citenamefont {Ran}\ \emph {et~al.}(2019)\citenamefont {Ran},
		\citenamefont {Eckberg}, \citenamefont {Ding}, \citenamefont {Furukawa},
		\citenamefont {Metz}, \citenamefont {Saha}, \citenamefont {Liu},
		\citenamefont {Zic}, \citenamefont {Kim}, \citenamefont {Paglione},\ and\
		\citenamefont {Butch}}]{Ran2019_UTe2}%
	\BibitemOpen
	\bibfield  {author} {\bibinfo {author} {\bibfnamefont {S.}~\bibnamefont
			{Ran}}, \bibinfo {author} {\bibfnamefont {C.}~\bibnamefont {Eckberg}},
		\bibinfo {author} {\bibfnamefont {Q.-P.}\ \bibnamefont {Ding}}, \bibinfo
		{author} {\bibfnamefont {Y.}~\bibnamefont {Furukawa}}, \bibinfo {author}
		{\bibfnamefont {T.}~\bibnamefont {Metz}}, \bibinfo {author} {\bibfnamefont
			{S.~R.}\ \bibnamefont {Saha}}, \bibinfo {author} {\bibfnamefont {I.-L.}\
			\bibnamefont {Liu}}, \bibinfo {author} {\bibfnamefont {M.}~\bibnamefont
			{Zic}}, \bibinfo {author} {\bibfnamefont {H.}~\bibnamefont {Kim}}, \bibinfo
		{author} {\bibfnamefont {J.}~\bibnamefont {Paglione}},\ and\ \bibinfo
		{author} {\bibfnamefont {N.~P.}\ \bibnamefont {Butch}},\ }\bibfield  {title}
	{\bibinfo {title} {Nearly ferromagnetic spin-triplet superconductivity},\
	}\href {https://doi.org/10.1126/science.aav8645} {\bibfield  {journal}
		{\bibinfo  {journal} {Science}\ }\textbf {\bibinfo {volume} {365}},\ \bibinfo
		{pages} {684–687} (\bibinfo {year} {2019})}\BibitemShut {NoStop}%
	\bibitem [{\citenamefont {Aoki}\ \emph {et~al.}(2019)\citenamefont {Aoki},
		\citenamefont {Nakamura}, \citenamefont {Honda}, \citenamefont {Li},
		\citenamefont {Homma}, \citenamefont {Shimizu}, \citenamefont {Sato},
		\citenamefont {Knebel}, \citenamefont {Brison}, \citenamefont {Pourret},
		\citenamefont {Braithwaite}, \citenamefont {Lapertot}, \citenamefont {Niu},
		\citenamefont {Vališka}, \citenamefont {Harima},\ and\ \citenamefont
		{Flouquet}}]{Aoki2019_UTe2}%
	\BibitemOpen
	\bibfield  {author} {\bibinfo {author} {\bibfnamefont {D.}~\bibnamefont
			{Aoki}}, \bibinfo {author} {\bibfnamefont {A.}~\bibnamefont {Nakamura}},
		\bibinfo {author} {\bibfnamefont {F.}~\bibnamefont {Honda}}, \bibinfo
		{author} {\bibfnamefont {D.}~\bibnamefont {Li}}, \bibinfo {author}
		{\bibfnamefont {Y.}~\bibnamefont {Homma}}, \bibinfo {author} {\bibfnamefont
			{Y.}~\bibnamefont {Shimizu}}, \bibinfo {author} {\bibfnamefont {Y.~J.}\
			\bibnamefont {Sato}}, \bibinfo {author} {\bibfnamefont {G.}~\bibnamefont
			{Knebel}}, \bibinfo {author} {\bibfnamefont {J.-P.}\ \bibnamefont {Brison}},
		\bibinfo {author} {\bibfnamefont {A.}~\bibnamefont {Pourret}}, \bibinfo
		{author} {\bibfnamefont {D.}~\bibnamefont {Braithwaite}}, \bibinfo {author}
		{\bibfnamefont {G.}~\bibnamefont {Lapertot}}, \bibinfo {author}
		{\bibfnamefont {Q.}~\bibnamefont {Niu}}, \bibinfo {author} {\bibfnamefont
			{M.}~\bibnamefont {Vališka}}, \bibinfo {author} {\bibfnamefont
			{H.}~\bibnamefont {Harima}},\ and\ \bibinfo {author} {\bibfnamefont
			{J.}~\bibnamefont {Flouquet}},\ }\bibfield  {title} {\bibinfo {title}
		{Unconventional superconductivity in heavy fermion {UTe$_2$}},\ }\href
	{https://doi.org/10.7566/jpsj.88.043702} {\bibfield  {journal} {\bibinfo
			{journal} {J. Phys. Soc. Jpn.}\ }\textbf {\bibinfo {volume} {88}},\ \bibinfo
		{pages} {043702} (\bibinfo {year} {2019})}\BibitemShut {NoStop}%
	\bibitem [{\citenamefont {Lai}\ \emph {et~al.}(2017)\citenamefont {Lai},
		\citenamefont {Grefe}, \citenamefont {Paschen},\ and\ \citenamefont
		{Si}}]{WeylKondo}%
	\BibitemOpen
	\bibfield  {author} {\bibinfo {author} {\bibfnamefont {H.-H.}\ \bibnamefont
			{Lai}}, \bibinfo {author} {\bibfnamefont {S.~E.}\ \bibnamefont {Grefe}},
		\bibinfo {author} {\bibfnamefont {S.}~\bibnamefont {Paschen}},\ and\ \bibinfo
		{author} {\bibfnamefont {Q.}~\bibnamefont {Si}},\ }\bibfield  {title}
	{\bibinfo {title} {{Weyl–Kondo semimetal in heavy-fermion systems}},\
	}\href {https://doi.org/10.1073/pnas.1715851115} {\bibfield  {journal}
		{\bibinfo  {journal} {Proc. Natl. Acad. Sci. USA}\ }\textbf {\bibinfo
			{volume} {115}},\ \bibinfo {pages} {93–97} (\bibinfo {year}
		{2017})}\BibitemShut {NoStop}%
	\bibitem [{\citenamefont {Neupane}\ \emph {et~al.}(2013)\citenamefont
		{Neupane}, \citenamefont {Alidoust}, \citenamefont {Xu}, \citenamefont
		{Kondo}, \citenamefont {Ishida}, \citenamefont {Kim}, \citenamefont {Liu},
		\citenamefont {Belopolski}, \citenamefont {Jo}, \citenamefont {Chang},
		\citenamefont {Jeng}, \citenamefont {Durakiewicz}, \citenamefont {Balicas},
		\citenamefont {Lin}, \citenamefont {Bansil}, \citenamefont {Shin},
		\citenamefont {Fisk},\ and\ \citenamefont {Hasan}}]{Neupane_KondoSmB6}%
	\BibitemOpen
	\bibfield  {author} {\bibinfo {author} {\bibfnamefont {M.}~\bibnamefont
			{Neupane}}, \bibinfo {author} {\bibfnamefont {N.}~\bibnamefont {Alidoust}},
		\bibinfo {author} {\bibfnamefont {S.-Y.}\ \bibnamefont {Xu}}, \bibinfo
		{author} {\bibfnamefont {T.}~\bibnamefont {Kondo}}, \bibinfo {author}
		{\bibfnamefont {Y.}~\bibnamefont {Ishida}}, \bibinfo {author} {\bibfnamefont
			{D.~J.}\ \bibnamefont {Kim}}, \bibinfo {author} {\bibfnamefont
			{C.}~\bibnamefont {Liu}}, \bibinfo {author} {\bibfnamefont {I.}~\bibnamefont
			{Belopolski}}, \bibinfo {author} {\bibfnamefont {Y.~J.}\ \bibnamefont {Jo}},
		\bibinfo {author} {\bibfnamefont {T.-R.}\ \bibnamefont {Chang}}, \bibinfo
		{author} {\bibfnamefont {H.-T.}\ \bibnamefont {Jeng}}, \bibinfo {author}
		{\bibfnamefont {T.}~\bibnamefont {Durakiewicz}}, \bibinfo {author}
		{\bibfnamefont {L.}~\bibnamefont {Balicas}}, \bibinfo {author} {\bibfnamefont
			{H.}~\bibnamefont {Lin}}, \bibinfo {author} {\bibfnamefont {A.}~\bibnamefont
			{Bansil}}, \bibinfo {author} {\bibfnamefont {S.}~\bibnamefont {Shin}},
		\bibinfo {author} {\bibfnamefont {Z.}~\bibnamefont {Fisk}},\ and\ \bibinfo
		{author} {\bibfnamefont {M.~Z.}\ \bibnamefont {Hasan}},\ }\bibfield  {title}
	{\bibinfo {title} {Surface electronic structure of the topological
			kondo-insulator candidate correlated electron system {SmB$_6$}},\ }\href
	{https://doi.org/10.1038/ncomms3991} {\bibfield  {journal} {\bibinfo
			{journal} {Nat. Commun.}\ }\textbf {\bibinfo {volume} {4}},\ \bibinfo {pages}
		{2991} (\bibinfo {year} {2013})}\BibitemShut {NoStop}%
	\bibitem [{\citenamefont {Khim}\ \emph {et~al.}(2021)\citenamefont {Khim},
		\citenamefont {Landaeta}, \citenamefont {Banda}, \citenamefont {Bannor},
		\citenamefont {Brando}, \citenamefont {Brydon}, \citenamefont {Hafner},
		\citenamefont {K\"{u}chler}, \citenamefont {Cardoso-Gil}, \citenamefont
		{Stockert}, \citenamefont {Mackenzie}, \citenamefont {Agterberg},
		\citenamefont {Geibel},\ and\ \citenamefont
		{Hassinger}}]{Superconductivity_CeRh2As2}%
	\BibitemOpen
	\bibfield  {author} {\bibinfo {author} {\bibfnamefont {S.}~\bibnamefont
			{Khim}}, \bibinfo {author} {\bibfnamefont {J.~F.}\ \bibnamefont {Landaeta}},
		\bibinfo {author} {\bibfnamefont {J.}~\bibnamefont {Banda}}, \bibinfo
		{author} {\bibfnamefont {N.}~\bibnamefont {Bannor}}, \bibinfo {author}
		{\bibfnamefont {M.}~\bibnamefont {Brando}}, \bibinfo {author} {\bibfnamefont
			{P.~M.~R.}\ \bibnamefont {Brydon}}, \bibinfo {author} {\bibfnamefont
			{D.}~\bibnamefont {Hafner}}, \bibinfo {author} {\bibfnamefont
			{R.}~\bibnamefont {K\"{u}chler}}, \bibinfo {author} {\bibfnamefont
			{R.}~\bibnamefont {Cardoso-Gil}}, \bibinfo {author} {\bibfnamefont
			{U.}~\bibnamefont {Stockert}}, \bibinfo {author} {\bibfnamefont {A.~P.}\
			\bibnamefont {Mackenzie}}, \bibinfo {author} {\bibfnamefont {D.~F.}\
			\bibnamefont {Agterberg}}, \bibinfo {author} {\bibfnamefont {C.}~\bibnamefont
			{Geibel}},\ and\ \bibinfo {author} {\bibfnamefont {E.}~\bibnamefont
			{Hassinger}},\ }\bibfield  {title} {\bibinfo {title} {Field-induced
			transition within the superconducting state of {CeRh$_2$As$_2$}},\ }\href
	{https://doi.org/10.1126/science.abe7518} {\bibfield  {journal} {\bibinfo
			{journal} {Science}\ }\textbf {\bibinfo {volume} {373}},\ \bibinfo {pages}
		{1012–1016} (\bibinfo {year} {2021})}\BibitemShut {NoStop}%
	\bibitem [{\citenamefont {M\"ockli}\ and\ \citenamefont
		{Ramires}(2021)}]{PRR_CeRh2As2}%
	\BibitemOpen
	\bibfield  {author} {\bibinfo {author} {\bibfnamefont {D.}~\bibnamefont
			{M\"ockli}}\ and\ \bibinfo {author} {\bibfnamefont {A.}~\bibnamefont
			{Ramires}},\ }\bibfield  {title} {\bibinfo {title} {Two scenarios for
			superconductivity in {CeRh$_2$As$_2$}},\ }\href
	{https://doi.org/10.1103/PhysRevResearch.3.023204} {\bibfield  {journal}
		{\bibinfo  {journal} {Phys. Rev. Res.}\ }\textbf {\bibinfo {volume} {3}},\
		\bibinfo {pages} {023204} (\bibinfo {year} {2021})}\BibitemShut {NoStop}%
	\bibitem [{\citenamefont {Nogaki}\ \emph {et~al.}(2021)\citenamefont {Nogaki},
		\citenamefont {Daido}, \citenamefont {Ishizuka},\ and\ \citenamefont
		{Yanase}}]{PRR_2021}%
	\BibitemOpen
	\bibfield  {author} {\bibinfo {author} {\bibfnamefont {K.}~\bibnamefont
			{Nogaki}}, \bibinfo {author} {\bibfnamefont {A.}~\bibnamefont {Daido}},
		\bibinfo {author} {\bibfnamefont {J.}~\bibnamefont {Ishizuka}},\ and\
		\bibinfo {author} {\bibfnamefont {Y.}~\bibnamefont {Yanase}},\ }\bibfield
	{title} {\bibinfo {title} {Topological crystalline superconductivity in
			locally noncentrosymmetric {CeRh$_2$As$_2$}},\ }\href
	{https://doi.org/10.1103/PhysRevResearch.3.L032071} {\bibfield  {journal}
		{\bibinfo  {journal} {Phys. Rev. Res.}\ }\textbf {\bibinfo {volume} {3}},\
		\bibinfo {pages} {L032071} (\bibinfo {year} {2021})}\BibitemShut {NoStop}%
	\bibitem [{\citenamefont {Zhu}\ \emph {et~al.}(2020{\natexlab{a}})\citenamefont
		{Zhu}, \citenamefont {Singh}, \citenamefont {Wang}, \citenamefont {Huang},
		\citenamefont {Chiu}, \citenamefont {Wang}, \citenamefont {Graf},
		\citenamefont {Zhang}, \citenamefont {Lin}, \citenamefont {Sun},
		\citenamefont {Bansil},\ and\ \citenamefont {Mao}}]{PRB_2020}%
	\BibitemOpen
	\bibfield  {author} {\bibinfo {author} {\bibfnamefont {Y.}~\bibnamefont
			{Zhu}}, \bibinfo {author} {\bibfnamefont {B.}~\bibnamefont {Singh}}, \bibinfo
		{author} {\bibfnamefont {Y.}~\bibnamefont {Wang}}, \bibinfo {author}
		{\bibfnamefont {C.-Y.}\ \bibnamefont {Huang}}, \bibinfo {author}
		{\bibfnamefont {W.-C.}\ \bibnamefont {Chiu}}, \bibinfo {author}
		{\bibfnamefont {B.}~\bibnamefont {Wang}}, \bibinfo {author} {\bibfnamefont
			{D.}~\bibnamefont {Graf}}, \bibinfo {author} {\bibfnamefont {Y.}~\bibnamefont
			{Zhang}}, \bibinfo {author} {\bibfnamefont {H.}~\bibnamefont {Lin}}, \bibinfo
		{author} {\bibfnamefont {J.}~\bibnamefont {Sun}}, \bibinfo {author}
		{\bibfnamefont {A.}~\bibnamefont {Bansil}},\ and\ \bibinfo {author}
		{\bibfnamefont {Z.}~\bibnamefont {Mao}},\ }\bibfield  {title} {\bibinfo
		{title} {{Exceptionally large anomalous Hall effect due to anticrossing of
				spin-split bands in the antiferromagnetic half-Heusler compound TbPtBi}},\
	}\href {https://doi.org/10.1103/PhysRevB.101.161105} {\bibfield  {journal}
		{\bibinfo  {journal} {Phys. Rev. B}\ }\textbf {\bibinfo {volume} {101}},\
		\bibinfo {pages} {161105} (\bibinfo {year} {2020}{\natexlab{a}})}\BibitemShut
	{NoStop}%
	\bibitem [{\citenamefont {Zhu}\ \emph {et~al.}(2023)\citenamefont {Zhu},
		\citenamefont {Huang}, \citenamefont {Wang}, \citenamefont {Graf},
		\citenamefont {Lin}, \citenamefont {Lee}, \citenamefont {Singleton},
		\citenamefont {Min}, \citenamefont {Palmstrom}, \citenamefont {Bansil},
		\citenamefont {Singh},\ and\ \citenamefont {Mao}}]{Zhu2023}%
	\BibitemOpen
	\bibfield  {author} {\bibinfo {author} {\bibfnamefont {Y.}~\bibnamefont
			{Zhu}}, \bibinfo {author} {\bibfnamefont {C.-Y.}\ \bibnamefont {Huang}},
		\bibinfo {author} {\bibfnamefont {Y.}~\bibnamefont {Wang}}, \bibinfo {author}
		{\bibfnamefont {D.}~\bibnamefont {Graf}}, \bibinfo {author} {\bibfnamefont
			{H.}~\bibnamefont {Lin}}, \bibinfo {author} {\bibfnamefont {S.~H.}\
			\bibnamefont {Lee}}, \bibinfo {author} {\bibfnamefont {J.}~\bibnamefont
			{Singleton}}, \bibinfo {author} {\bibfnamefont {L.}~\bibnamefont {Min}},
		\bibinfo {author} {\bibfnamefont {J.~C.}\ \bibnamefont {Palmstrom}}, \bibinfo
		{author} {\bibfnamefont {A.}~\bibnamefont {Bansil}}, \bibinfo {author}
		{\bibfnamefont {B.}~\bibnamefont {Singh}},\ and\ \bibinfo {author}
		{\bibfnamefont {Z.}~\bibnamefont {Mao}},\ }\bibfield  {title} {\bibinfo
		{title} {Large anomalous hall effect and negative magnetoresistance in
			half-topological semimetals},\ }\href
	{https://doi.org/10.1038/s42005-023-01469-6} {\bibfield  {journal} {\bibinfo
			{journal} {Commun. phys.}\ }\textbf {\bibinfo {volume} {6}},\ \bibinfo
		{pages} {346} (\bibinfo {year} {2023})}\BibitemShut {NoStop}%
	\bibitem [{\citenamefont {Nandi}\ \emph {et~al.}(2020)\citenamefont {Nandi},
		\citenamefont {Thamizhavel},\ and\ \citenamefont {Dhar}}]{Nandi_2020}%
	\BibitemOpen
	\bibfield  {author} {\bibinfo {author} {\bibfnamefont {M.}~\bibnamefont
			{Nandi}}, \bibinfo {author} {\bibfnamefont {A.}~\bibnamefont {Thamizhavel}},\
		and\ \bibinfo {author} {\bibfnamefont {S.~K.}\ \bibnamefont {Dhar}},\
	}\bibfield  {title} {\bibinfo {title} {{Anisotropic magnetic properties of
				trigonal ErAl$_2$Ge$_2$ single crystal}},\ }\href
	{https://doi.org/10.1088/1361-648X/ab6d16} {\bibfield  {journal} {\bibinfo
			{journal} {J. Phys.: Condens. Matter}\ }\textbf {\bibinfo {volume} {32}},\
		\bibinfo {pages} {185803} (\bibinfo {year} {2020})}\BibitemShut {NoStop}%
	\bibitem [{\citenamefont {Khan}\ \emph {et~al.}(2024)\citenamefont {Khan},
		\citenamefont {Iqbal}, \citenamefont {Rehman}, \citenamefont {Alam},
		\citenamefont {Rahim}, \citenamefont {Hussain}, \citenamefont {{Nawaz
				Sharif}},\ and\ \citenamefont {{Ajmal Khan}}}]{KHAN}%
	\BibitemOpen
	\bibfield  {author} {\bibinfo {author} {\bibfnamefont {J.}~\bibnamefont
			{Khan}}, \bibinfo {author} {\bibfnamefont {Y.}~\bibnamefont {Iqbal}},
		\bibinfo {author} {\bibfnamefont {F.}~\bibnamefont {Rehman}}, \bibinfo
		{author} {\bibfnamefont {J.}~\bibnamefont {Alam}}, \bibinfo {author}
		{\bibfnamefont {M.}~\bibnamefont {Rahim}}, \bibinfo {author} {\bibfnamefont
			{S.}~\bibnamefont {Hussain}}, \bibinfo {author} {\bibfnamefont
			{M.}~\bibnamefont {{Nawaz Sharif}}},\ and\ \bibinfo {author} {\bibfnamefont
			{M.}~\bibnamefont {{Ajmal Khan}}},\ }\bibfield  {title} {\bibinfo {title}
		{{Exploring Rare-Earth compound ErMn$_2$Si$_2$ for thermoelectric and
				photoelectrochemical applications using Density functional theory}},\ }\href
	{https://doi.org/https://doi.org/10.1016/j.jmmm.2024.172156} {\bibfield
		{journal} {\bibinfo  {journal} {J. Magn. Magn. Mater.}\ }\textbf {\bibinfo
			{volume} {600}},\ \bibinfo {pages} {172156} (\bibinfo {year}
		{2024})}\BibitemShut {NoStop}%
	\bibitem [{\citenamefont {Venturini}\ \emph {et~al.}(1995)\citenamefont
		{Venturini}, \citenamefont {Welter}, \citenamefont {Ressouche},\ and\
		\citenamefont {Malaman}}]{Neutron1995}%
	\BibitemOpen
	\bibfield  {author} {\bibinfo {author} {\bibfnamefont {G.}~\bibnamefont
			{Venturini}}, \bibinfo {author} {\bibfnamefont {R.}~\bibnamefont {Welter}},
		\bibinfo {author} {\bibfnamefont {E.}~\bibnamefont {Ressouche}},\ and\
		\bibinfo {author} {\bibfnamefont {B.}~\bibnamefont {Malaman}},\ }\bibfield
	{title} {\bibinfo {title} {Neutron diffraction study of
			{Nd$_0.35$La$_0.65$Mn$_2$Si$_2$}: A {SmMn$_2$Ge$_2$}-like magnetic behaviour
			compound},\ }\href {https://doi.org/10.1016/0304-8853(95)00120-4} {\bibfield
		{journal} {\bibinfo  {journal} {J. Magn. Magn. Mater.}\ }\textbf {\bibinfo
			{volume} {150}},\ \bibinfo {pages} {197–212} (\bibinfo {year}
		{1995})}\BibitemShut {NoStop}%
	\bibitem [{\citenamefont {Barla}\ \emph {et~al.}(2004)\citenamefont {Barla},
		\citenamefont {Sanchez}, \citenamefont {Malaman}, \citenamefont {Doyle},\
		and\ \citenamefont {R\"uffer}}]{PRB2004}%
	\BibitemOpen
	\bibfield  {author} {\bibinfo {author} {\bibfnamefont {A.}~\bibnamefont
			{Barla}}, \bibinfo {author} {\bibfnamefont {J.~P.}\ \bibnamefont {Sanchez}},
		\bibinfo {author} {\bibfnamefont {B.}~\bibnamefont {Malaman}}, \bibinfo
		{author} {\bibfnamefont {B.~P.}\ \bibnamefont {Doyle}},\ and\ \bibinfo
		{author} {\bibfnamefont {R.}~\bibnamefont {R\"uffer}},\ }\bibfield  {title}
	{\bibinfo {title} {{Sm magnetism in the layered compound
				SmMn$_{2}$Ge$_{2}$}},\ }\href {https://doi.org/10.1103/PhysRevB.69.220405}
	{\bibfield  {journal} {\bibinfo  {journal} {Phys. Rev. B}\ }\textbf {\bibinfo
			{volume} {69}},\ \bibinfo {pages} {220405} (\bibinfo {year}
		{2004})}\BibitemShut {NoStop}%
	\bibitem [{\citenamefont {Brabers}\ \emph {et~al.}(1993)\citenamefont
		{Brabers}, \citenamefont {Bakker}, \citenamefont {Nakotte}, \citenamefont
		{de~Boer}, \citenamefont {Lenczowski},\ and\ \citenamefont
		{Buschow}}]{Brabers1993}%
	\BibitemOpen
	\bibfield  {author} {\bibinfo {author} {\bibfnamefont {J.}~\bibnamefont
			{Brabers}}, \bibinfo {author} {\bibfnamefont {K.}~\bibnamefont {Bakker}},
		\bibinfo {author} {\bibfnamefont {H.}~\bibnamefont {Nakotte}}, \bibinfo
		{author} {\bibfnamefont {F.}~\bibnamefont {de~Boer}}, \bibinfo {author}
		{\bibfnamefont {S.}~\bibnamefont {Lenczowski}},\ and\ \bibinfo {author}
		{\bibfnamefont {K.}~\bibnamefont {Buschow}},\ }\bibfield  {title} {\bibinfo
		{title} {Giant magnetoresistance in polycrystalline {SmMn$_2$Ge$_2$}},\
	}\href {https://doi.org/10.1016/0925-8388(93)90419-n} {\bibfield  {journal}
		{\bibinfo  {journal} {J. Alloys Compd.}\ }\textbf {\bibinfo {volume} {199}},\
		\bibinfo {pages} {L1–L3} (\bibinfo {year} {1993})}\BibitemShut {NoStop}%
	\bibitem [{\citenamefont {Brabers}\ \emph {et~al.}(1994)\citenamefont
		{Brabers}, \citenamefont {Nolten}, \citenamefont {Kayzel}, \citenamefont
		{Lenczowski}, \citenamefont {Buschow},\ and\ \citenamefont
		{de~Boer}}]{PRB1994}%
	\BibitemOpen
	\bibfield  {author} {\bibinfo {author} {\bibfnamefont {J.~H. V.~J.}\
			\bibnamefont {Brabers}}, \bibinfo {author} {\bibfnamefont {A.~J.}\
			\bibnamefont {Nolten}}, \bibinfo {author} {\bibfnamefont {F.}~\bibnamefont
			{Kayzel}}, \bibinfo {author} {\bibfnamefont {S.~H.~J.}\ \bibnamefont
			{Lenczowski}}, \bibinfo {author} {\bibfnamefont {K.~H.~J.}\ \bibnamefont
			{Buschow}},\ and\ \bibinfo {author} {\bibfnamefont {F.~R.}\ \bibnamefont
			{de~Boer}},\ }\bibfield  {title} {\bibinfo {title} {{Strong Mn-Mn distance
				dependence of the Mn interlayer coupling in SmMn$_{2}$Ge$_{2}$-related
				compounds and its role in magnetic phase transitions}},\ }\href
	{https://doi.org/10.1103/PhysRevB.50.16410} {\bibfield  {journal} {\bibinfo
			{journal} {Phys. Rev. B}\ }\textbf {\bibinfo {volume} {50}},\ \bibinfo
		{pages} {16410} (\bibinfo {year} {1994})}\BibitemShut {NoStop}%
	\bibitem [{\citenamefont {Singh}\ \emph {et~al.}(2024)\citenamefont {Singh},
		\citenamefont {Sau}, \citenamefont {Rai}, \citenamefont {Panda},
		\citenamefont {Kumar},\ and\ \citenamefont {Kumar}}]{PRM_2024}%
	\BibitemOpen
	\bibfield  {author} {\bibinfo {author} {\bibfnamefont {M.}~\bibnamefont
			{Singh}}, \bibinfo {author} {\bibfnamefont {J.}~\bibnamefont {Sau}}, \bibinfo
		{author} {\bibfnamefont {B.}~\bibnamefont {Rai}}, \bibinfo {author}
		{\bibfnamefont {A.}~\bibnamefont {Panda}}, \bibinfo {author} {\bibfnamefont
			{M.}~\bibnamefont {Kumar}},\ and\ \bibinfo {author} {\bibfnamefont
			{N.}~\bibnamefont {Kumar}},\ }\bibfield  {title} {\bibinfo {title} {Tuning
			intrinsic anomalous hall effect from large to zero in two ferromagnetic
			states of {SmMn}$_{2}${Ge}$_{2}$},\ }\href
	{https://doi.org/10.1103/PhysRevMaterials.8.084201} {\bibfield  {journal}
		{\bibinfo  {journal} {Phys. Rev. Mater.}\ }\textbf {\bibinfo {volume} {8}},\
		\bibinfo {pages} {084201} (\bibinfo {year} {2024})}\BibitemShut {NoStop}%
	\bibitem [{\citenamefont {Joshi}\ \emph {et~al.}(2009)\citenamefont {Joshi},
		\citenamefont {Nagalakshmi}, \citenamefont {Kulkarni}, \citenamefont {Dhar},\
		and\ \citenamefont {Thamizhavel}}]{JOSHI}%
	\BibitemOpen
	\bibfield  {author} {\bibinfo {author} {\bibfnamefont {D.~A.}\ \bibnamefont
			{Joshi}}, \bibinfo {author} {\bibfnamefont {R.}~\bibnamefont {Nagalakshmi}},
		\bibinfo {author} {\bibfnamefont {R.}~\bibnamefont {Kulkarni}}, \bibinfo
		{author} {\bibfnamefont {S.}~\bibnamefont {Dhar}},\ and\ \bibinfo {author}
		{\bibfnamefont {A.}~\bibnamefont {Thamizhavel}},\ }\bibfield  {title}
	{\bibinfo {title} {{Crystal growth and anisotropic magnetic properties of
				RAg$_2$Ge$_2$ (R=Pr, Nd and Sm) single crystals}},\ }\href
	{https://doi.org/https://doi.org/10.1016/j.physb.2009.07.016} {\bibfield
		{journal} {\bibinfo  {journal} {Physica B: Condensed Matter}\ }\textbf
		{\bibinfo {volume} {404}},\ \bibinfo {pages} {2988} (\bibinfo {year}
		{2009})}\BibitemShut {NoStop}%
	\bibitem [{\citenamefont
		{Rodríguez-Carvajal}(1993)}]{RODRIGUEZCARVAJAL199355}%
	\BibitemOpen
	\bibfield  {author} {\bibinfo {author} {\bibfnamefont {J.}~\bibnamefont
			{Rodríguez-Carvajal}},\ }\bibfield  {title} {\bibinfo {title} {Recent
			advances in magnetic structure determination by neutron powder diffraction},\
	}\href {https://doi.org/https://doi.org/10.1016/0921-4526(93)90108-I}
	{\bibfield  {journal} {\bibinfo  {journal} {Physica B: Condensed Matter}\
		}\textbf {\bibinfo {volume} {192}},\ \bibinfo {pages} {55} (\bibinfo {year}
		{1993})}\BibitemShut {NoStop}%
	\bibitem [{\citenamefont {Hohenberg}\ and\ \citenamefont {Kohn}(1964)}]{Hohen}%
	\BibitemOpen
	\bibfield  {author} {\bibinfo {author} {\bibfnamefont {P.}~\bibnamefont
			{Hohenberg}}\ and\ \bibinfo {author} {\bibfnamefont {W.}~\bibnamefont
			{Kohn}},\ }\bibfield  {title} {\bibinfo {title} {Inhomogeneous electron
			gas},\ }\href {https://doi.org/10.1103/PhysRev.136.B864} {\bibfield
		{journal} {\bibinfo  {journal} {Phys. Rev.}\ }\textbf {\bibinfo {volume}
			{136}},\ \bibinfo {pages} {B864} (\bibinfo {year} {1964})}\BibitemShut
	{NoStop}%
	\bibitem [{\citenamefont {Bl\"ochl}(1994)}]{Bloch}%
	\BibitemOpen
	\bibfield  {author} {\bibinfo {author} {\bibfnamefont {P.~E.}\ \bibnamefont
			{Bl\"ochl}},\ }\bibfield  {title} {\bibinfo {title} {Projector augmented-wave
			method},\ }\href {https://doi.org/10.1103/PhysRevB.50.17953} {\bibfield
		{journal} {\bibinfo  {journal} {Phys. Rev. B}\ }\textbf {\bibinfo {volume}
			{50}},\ \bibinfo {pages} {17953} (\bibinfo {year} {1994})}\BibitemShut
	{NoStop}%
	\bibitem [{\citenamefont {Kresse}\ and\ \citenamefont
		{Furthm\"uller}(1996)}]{Kresse1996}%
	\BibitemOpen
	\bibfield  {author} {\bibinfo {author} {\bibfnamefont {G.}~\bibnamefont
			{Kresse}}\ and\ \bibinfo {author} {\bibfnamefont {J.}~\bibnamefont
			{Furthm\"uller}},\ }\bibfield  {title} {\bibinfo {title} {Efficient iterative
			schemes for $ab ~initio$ total-energy calculations using a plane-wave basis
			set},\ }\href {https://doi.org/10.1103/PhysRevB.54.11169} {\bibfield
		{journal} {\bibinfo  {journal} {Phys. Rev. B}\ }\textbf {\bibinfo {volume}
			{54}},\ \bibinfo {pages} {11169} (\bibinfo {year} {1996})}\BibitemShut
	{NoStop}%
	\bibitem [{\citenamefont {Kresse}\ and\ \citenamefont
		{Joubert}(1999)}]{Kresse1999}%
	\BibitemOpen
	\bibfield  {author} {\bibinfo {author} {\bibfnamefont {G.}~\bibnamefont
			{Kresse}}\ and\ \bibinfo {author} {\bibfnamefont {D.}~\bibnamefont
			{Joubert}},\ }\bibfield  {title} {\bibinfo {title} {From ultrasoft
			pseudopotentials to the projector augmented-wave method},\ }\href
	{https://doi.org/10.1103/PhysRevB.59.1758} {\bibfield  {journal} {\bibinfo
			{journal} {Phys. Rev. B}\ }\textbf {\bibinfo {volume} {59}},\ \bibinfo
		{pages} {1758} (\bibinfo {year} {1999})}\BibitemShut {NoStop}%
	\bibitem [{\citenamefont {Perdew}\ \emph {et~al.}(1996)\citenamefont {Perdew},
		\citenamefont {Burke},\ and\ \citenamefont
		{Ernzerhof}}]{perdew1996generalized}%
	\BibitemOpen
	\bibfield  {author} {\bibinfo {author} {\bibfnamefont {J.~P.}\ \bibnamefont
			{Perdew}}, \bibinfo {author} {\bibfnamefont {K.}~\bibnamefont {Burke}},\ and\
		\bibinfo {author} {\bibfnamefont {M.}~\bibnamefont {Ernzerhof}},\ }\bibfield
	{title} {\bibinfo {title} {Generalized gradient approximation made simple},\
	}\href {https://doi.org/10.1103/PhysRevLett.77.3865} {\bibfield  {journal}
		{\bibinfo  {journal} {Phys. Rev. Lett.}\ }\textbf {\bibinfo {volume} {77}},\
		\bibinfo {pages} {3865} (\bibinfo {year} {1996})}\BibitemShut {NoStop}%
	\bibitem [{\citenamefont {Anisimov}\ \emph {et~al.}(1991)\citenamefont
		{Anisimov}, \citenamefont {Zaanen},\ and\ \citenamefont
		{Andersen}}]{HubbardU}%
	\BibitemOpen
	\bibfield  {author} {\bibinfo {author} {\bibfnamefont {V.~I.}\ \bibnamefont
			{Anisimov}}, \bibinfo {author} {\bibfnamefont {J.}~\bibnamefont {Zaanen}},\
		and\ \bibinfo {author} {\bibfnamefont {O.~K.}\ \bibnamefont {Andersen}},\
	}\bibfield  {title} {\bibinfo {title} {Band theory and mott insulators:
			Hubbard {U} instead of stoner {I}},\ }\href
	{https://doi.org/10.1103/PhysRevB.44.943} {\bibfield  {journal} {\bibinfo
			{journal} {Phys. Rev. B}\ }\textbf {\bibinfo {volume} {44}},\ \bibinfo
		{pages} {943} (\bibinfo {year} {1991})}\BibitemShut {NoStop}%
	\bibitem [{\citenamefont {Anisimov}\ \emph {et~al.}(1997)\citenamefont
		{Anisimov}, \citenamefont {Poteryaev}, \citenamefont {Korotin}, \citenamefont
		{Anokhin},\ and\ \citenamefont {Kotliar}}]{LDAU}%
	\BibitemOpen
	\bibfield  {author} {\bibinfo {author} {\bibfnamefont {V.~I.}\ \bibnamefont
			{Anisimov}}, \bibinfo {author} {\bibfnamefont {A.~I.}\ \bibnamefont
			{Poteryaev}}, \bibinfo {author} {\bibfnamefont {M.~A.}\ \bibnamefont
			{Korotin}}, \bibinfo {author} {\bibfnamefont {A.~O.}\ \bibnamefont
			{Anokhin}},\ and\ \bibinfo {author} {\bibfnamefont {G.}~\bibnamefont
			{Kotliar}},\ }\bibfield  {title} {\bibinfo {title} {First-principles
			calculations of the electronic structure and spectra of strongly correlated
			systems: dynamical mean-field theory},\ }\href
	{https://doi.org/10.1088/0953-8984/9/35/010} {\bibfield  {journal} {\bibinfo
			{journal} {J. Phys.: Condens. Matter}\ }\textbf {\bibinfo {volume} {9}},\
		\bibinfo {pages} {7359–7367} (\bibinfo {year} {1997})}\BibitemShut
	{NoStop}%
	\bibitem [{\citenamefont {Mostofi}\ \emph {et~al.}(2008)\citenamefont
		{Mostofi}, \citenamefont {Yates}, \citenamefont {Lee}, \citenamefont {Souza},
		\citenamefont {Vanderbilt},\ and\ \citenamefont
		{Marzari}}]{mostofi2008wannier90}%
	\BibitemOpen
	\bibfield  {author} {\bibinfo {author} {\bibfnamefont {A.~A.}\ \bibnamefont
			{Mostofi}}, \bibinfo {author} {\bibfnamefont {J.~R.}\ \bibnamefont {Yates}},
		\bibinfo {author} {\bibfnamefont {Y.-S.}\ \bibnamefont {Lee}}, \bibinfo
		{author} {\bibfnamefont {I.}~\bibnamefont {Souza}}, \bibinfo {author}
		{\bibfnamefont {D.}~\bibnamefont {Vanderbilt}},\ and\ \bibinfo {author}
		{\bibfnamefont {N.}~\bibnamefont {Marzari}},\ }\bibfield  {title} {\bibinfo
		{title} {Wannier90: A tool for obtaining maximally-localised wannier
			functions},\ }\href
	{https://doi.org/https://doi.org/10.1016/j.cpc.2007.11.016} {\bibfield
		{journal} {\bibinfo  {journal} {Comput. Phys. Commun.}\ }\textbf {\bibinfo
			{volume} {178}},\ \bibinfo {pages} {685} (\bibinfo {year}
		{2008})}\BibitemShut {NoStop}%
	\bibitem [{\citenamefont {Wu}\ \emph {et~al.}(2018)\citenamefont {Wu},
		\citenamefont {Zhang}, \citenamefont {Song}, \citenamefont {Troyer},\ and\
		\citenamefont {Soluyanov}}]{Wtools}%
	\BibitemOpen
	\bibfield  {author} {\bibinfo {author} {\bibfnamefont {Q.}~\bibnamefont
			{Wu}}, \bibinfo {author} {\bibfnamefont {S.}~\bibnamefont {Zhang}}, \bibinfo
		{author} {\bibfnamefont {H.-F.}\ \bibnamefont {Song}}, \bibinfo {author}
		{\bibfnamefont {M.}~\bibnamefont {Troyer}},\ and\ \bibinfo {author}
		{\bibfnamefont {A.~A.}\ \bibnamefont {Soluyanov}},\ }\bibfield  {title}
	{\bibinfo {title} {Wannier{T}ools : An open-source software package for novel
			topological materials},\ }\href
	{https://doi.org/https://doi.org/10.1016/j.cpc.2017.09.033} {\bibfield
		{journal} {\bibinfo  {journal} {Comput. Phys. Commun.}\ }\textbf {\bibinfo
			{volume} {224}},\ \bibinfo {pages} {405 } (\bibinfo {year}
		{2018})}\BibitemShut {NoStop}%
	\bibitem [{\citenamefont {Rourke}\ and\ \citenamefont {Julian}(2012)}]{skeaf}%
	\BibitemOpen
	\bibfield  {author} {\bibinfo {author} {\bibfnamefont {P.}~\bibnamefont
			{Rourke}}\ and\ \bibinfo {author} {\bibfnamefont {S.}~\bibnamefont
			{Julian}},\ }\bibfield  {title} {\bibinfo {title} {{Numerical extraction of
				de Haas–van Alphen frequencies from calculated band energies}},\ }\href
	{https://doi.org/https://doi.org/10.1016/j.cpc.2011.10.015} {\bibfield
		{journal} {\bibinfo  {journal} {Comput. Phys. Commun.}\ }\textbf {\bibinfo
			{volume} {183}},\ \bibinfo {pages} {324} (\bibinfo {year}
		{2012})}\BibitemShut {NoStop}%
	\bibitem [{\citenamefont {Anand}\ \emph {et~al.}(2014)\citenamefont {Anand},
		\citenamefont {Kim}, \citenamefont {Tanatar}, \citenamefont {Prozorov},\ and\
		\citenamefont {Johnston}}]{Anand2014}%
	\BibitemOpen
	\bibfield  {author} {\bibinfo {author} {\bibfnamefont {V.~K.}\ \bibnamefont
			{Anand}}, \bibinfo {author} {\bibfnamefont {H.}~\bibnamefont {Kim}}, \bibinfo
		{author} {\bibfnamefont {M.~A.}\ \bibnamefont {Tanatar}}, \bibinfo {author}
		{\bibfnamefont {R.}~\bibnamefont {Prozorov}},\ and\ \bibinfo {author}
		{\bibfnamefont {D.~C.}\ \bibnamefont {Johnston}},\ }\bibfield  {title}
	{\bibinfo {title} {{Superconductivity and physical properties of
				CaPd$_2$Ge$_2$ single crystals}},\ }\href
	{https://doi.org/10.1088/0953-8984/26/40/405702} {\bibfield  {journal}
		{\bibinfo  {journal} {J. Phys.: Condens. Matter}\ }\textbf {\bibinfo {volume}
			{26}},\ \bibinfo {pages} {405702} (\bibinfo {year} {2014})}\BibitemShut
	{NoStop}%
	\bibitem [{\citenamefont {Midya}\ \emph {et~al.}(2016)\citenamefont {Midya},
		\citenamefont {Mandal}, \citenamefont {Rubi}, \citenamefont {Chen},
		\citenamefont {Wang}, \citenamefont {Mahendiran}, \citenamefont {Lorusso},\
		and\ \citenamefont {Evangelisti}}]{PhysRevB.93.094422}%
	\BibitemOpen
	\bibfield  {author} {\bibinfo {author} {\bibfnamefont {A.}~\bibnamefont
			{Midya}}, \bibinfo {author} {\bibfnamefont {P.}~\bibnamefont {Mandal}},
		\bibinfo {author} {\bibfnamefont {K.}~\bibnamefont {Rubi}}, \bibinfo {author}
		{\bibfnamefont {R.}~\bibnamefont {Chen}}, \bibinfo {author} {\bibfnamefont
			{J.-S.}\ \bibnamefont {Wang}}, \bibinfo {author} {\bibfnamefont
			{R.}~\bibnamefont {Mahendiran}}, \bibinfo {author} {\bibfnamefont
			{G.}~\bibnamefont {Lorusso}},\ and\ \bibinfo {author} {\bibfnamefont
			{M.}~\bibnamefont {Evangelisti}},\ }\bibfield  {title} {\bibinfo {title}
		{{Large adiabatic temperature and magnetic entropy changes in EuTiO$_3$}},\
	}\href {https://doi.org/10.1103/PhysRevB.93.094422} {\bibfield  {journal}
		{\bibinfo  {journal} {Phys. Rev. B}\ }\textbf {\bibinfo {volume} {93}},\
		\bibinfo {pages} {094422} (\bibinfo {year} {2016})}\BibitemShut {NoStop}%
	\bibitem [{\citenamefont {Abrikosov}(2000)}]{AAAbrikosov_2000}%
	\BibitemOpen
	\bibfield  {author} {\bibinfo {author} {\bibfnamefont {A.~A.}\ \bibnamefont
			{Abrikosov}},\ }\bibfield  {title} {\bibinfo {title} {Quantum linear
			magnetoresistance},\ }\href {https://doi.org/10.1209/epl/i2000-00220-2}
	{\bibfield  {journal} {\bibinfo  {journal} {Europhysics Letters}\ }\textbf
		{\bibinfo {volume} {49}},\ \bibinfo {pages} {789} (\bibinfo {year}
		{2000})}\BibitemShut {NoStop}%
	\bibitem [{\citenamefont {Parish}\ and\ \citenamefont
		{Littlewood}(2003)}]{Parish2003}%
	\BibitemOpen
	\bibfield  {author} {\bibinfo {author} {\bibfnamefont {M.~M.}\ \bibnamefont
			{Parish}}\ and\ \bibinfo {author} {\bibfnamefont {P.~B.}\ \bibnamefont
			{Littlewood}},\ }\bibfield  {title} {\bibinfo {title} {Non-saturating
			magnetoresistance in heavily disordered semiconductors},\ }\href
	{https://doi.org/10.1038/nature02073} {\bibfield  {journal} {\bibinfo
			{journal} {Nature}\ }\textbf {\bibinfo {volume} {426}},\ \bibinfo {pages}
		{162} (\bibinfo {year} {2003})}\BibitemShut {NoStop}%
	\bibitem [{\citenamefont {Abrikosov}(1998)}]{PhysRevB.58.2788}%
	\BibitemOpen
	\bibfield  {author} {\bibinfo {author} {\bibfnamefont {A.~A.}\ \bibnamefont
			{Abrikosov}},\ }\bibfield  {title} {\bibinfo {title} {Quantum
			magnetoresistance},\ }\href {https://doi.org/10.1103/PhysRevB.58.2788}
	{\bibfield  {journal} {\bibinfo  {journal} {Phys. Rev. B}\ }\textbf {\bibinfo
			{volume} {58}},\ \bibinfo {pages} {2788} (\bibinfo {year}
		{1998})}\BibitemShut {NoStop}%
	\bibitem [{\citenamefont {Hu}\ \emph {et~al.}(2007)\citenamefont {Hu},
		\citenamefont {Parish},\ and\ \citenamefont
		{Rosenbaum}}]{PhysRevB.75.214203}%
	\BibitemOpen
	\bibfield  {author} {\bibinfo {author} {\bibfnamefont {J.}~\bibnamefont
			{Hu}}, \bibinfo {author} {\bibfnamefont {M.~M.}\ \bibnamefont {Parish}},\
		and\ \bibinfo {author} {\bibfnamefont {T.~F.}\ \bibnamefont {Rosenbaum}},\
	}\bibfield  {title} {\bibinfo {title} {Nonsaturating magnetoresistance of
			inhomogeneous conductors: Comparison of experiment and simulation},\ }\href
	{https://doi.org/10.1103/PhysRevB.75.214203} {\bibfield  {journal} {\bibinfo
			{journal} {Phys. Rev. B}\ }\textbf {\bibinfo {volume} {75}},\ \bibinfo
		{pages} {214203} (\bibinfo {year} {2007})}\BibitemShut {NoStop}%
	\bibitem [{\citenamefont {Feng}\ \emph {et~al.}(2015)\citenamefont {Feng},
		\citenamefont {Pang}, \citenamefont {Wu}, \citenamefont {Wang}, \citenamefont
		{Weng}, \citenamefont {Li}, \citenamefont {Dai}, \citenamefont {Fang},
		\citenamefont {Shi},\ and\ \citenamefont {Lu}}]{feng2015large}%
	\BibitemOpen
	\bibfield  {author} {\bibinfo {author} {\bibfnamefont {J.}~\bibnamefont
			{Feng}}, \bibinfo {author} {\bibfnamefont {Y.}~\bibnamefont {Pang}}, \bibinfo
		{author} {\bibfnamefont {D.}~\bibnamefont {Wu}}, \bibinfo {author}
		{\bibfnamefont {Z.}~\bibnamefont {Wang}}, \bibinfo {author} {\bibfnamefont
			{H.}~\bibnamefont {Weng}}, \bibinfo {author} {\bibfnamefont {J.}~\bibnamefont
			{Li}}, \bibinfo {author} {\bibfnamefont {X.}~\bibnamefont {Dai}}, \bibinfo
		{author} {\bibfnamefont {Z.}~\bibnamefont {Fang}}, \bibinfo {author}
		{\bibfnamefont {Y.}~\bibnamefont {Shi}},\ and\ \bibinfo {author}
		{\bibfnamefont {L.}~\bibnamefont {Lu}},\ }\bibfield  {title} {\bibinfo
		{title} {{Large linear magnetoresistance in Dirac semimetal Cd$_3$As$_2$ with
				Fermi surfaces close to the Dirac points}},\ }\href
	{https://doi.org/https://doi.org/10.1103/PhysRevB.92.081306} {\bibfield
		{journal} {\bibinfo  {journal} {Physical Review B}\ }\textbf {\bibinfo
			{volume} {92}},\ \bibinfo {pages} {081306} (\bibinfo {year}
		{2015})}\BibitemShut {NoStop}%
	\bibitem [{\citenamefont {Zhou}\ \emph {et~al.}(2020)\citenamefont {Zhou},
		\citenamefont {Lou}, \citenamefont {Zhang}, \citenamefont {Chen},
		\citenamefont {Chen}, \citenamefont {Xu}, \citenamefont {Du}, \citenamefont
		{Yang}, \citenamefont {Wang}, \citenamefont {Xi} \emph
		{et~al.}}]{zhou2020linear}%
	\BibitemOpen
	\bibfield  {author} {\bibinfo {author} {\bibfnamefont {Y.}~\bibnamefont
			{Zhou}}, \bibinfo {author} {\bibfnamefont {Z.}~\bibnamefont {Lou}}, \bibinfo
		{author} {\bibfnamefont {S.}~\bibnamefont {Zhang}}, \bibinfo {author}
		{\bibfnamefont {H.}~\bibnamefont {Chen}}, \bibinfo {author} {\bibfnamefont
			{Q.}~\bibnamefont {Chen}}, \bibinfo {author} {\bibfnamefont {B.}~\bibnamefont
			{Xu}}, \bibinfo {author} {\bibfnamefont {J.}~\bibnamefont {Du}}, \bibinfo
		{author} {\bibfnamefont {J.}~\bibnamefont {Yang}}, \bibinfo {author}
		{\bibfnamefont {H.}~\bibnamefont {Wang}}, \bibinfo {author} {\bibfnamefont
			{C.}~\bibnamefont {Xi}}, \emph {et~al.},\ }\bibfield  {title} {\bibinfo
		{title} {{Linear and quadratic magnetoresistance in the semimetal SiP$_2$}},\
	}\href {https://doi.org/https://doi.org/10.1103/PhysRevB.102.115145}
	{\bibfield  {journal} {\bibinfo  {journal} {Physical Review B}\ }\textbf
		{\bibinfo {volume} {102}},\ \bibinfo {pages} {115145} (\bibinfo {year}
		{2020})}\BibitemShut {NoStop}%
	\bibitem [{SM()}]{SM}%
	\BibitemOpen
	\bibfield  {title} {\bibinfo {title} {[url will be inserted by publisher]
			Estimation of anomalous Hall effect and simultaneous
			fitting of the two band model, which does not match well with the experimental data are shown in the Supplemental material}}\href@noop {} {\ }\BibitemShut {NoStop}%
	\bibitem [{\citenamefont {Shekhar}\ \emph {et~al.}(2018)\citenamefont
		{Shekhar}, \citenamefont {Kumar}, \citenamefont {Grinenko}, \citenamefont
		{Singh}, \citenamefont {Sarkar}, \citenamefont {Luetkens}, \citenamefont
		{Wu}, \citenamefont {Zhang}, \citenamefont {Komarek}, \citenamefont {Kampert}
		\emph {et~al.}}]{shekhar2018anomalous}%
	\BibitemOpen
	\bibfield  {author} {\bibinfo {author} {\bibfnamefont {C.}~\bibnamefont
			{Shekhar}}, \bibinfo {author} {\bibfnamefont {N.}~\bibnamefont {Kumar}},
		\bibinfo {author} {\bibfnamefont {V.}~\bibnamefont {Grinenko}}, \bibinfo
		{author} {\bibfnamefont {S.}~\bibnamefont {Singh}}, \bibinfo {author}
		{\bibfnamefont {R.}~\bibnamefont {Sarkar}}, \bibinfo {author} {\bibfnamefont
			{H.}~\bibnamefont {Luetkens}}, \bibinfo {author} {\bibfnamefont {S.-C.}\
			\bibnamefont {Wu}}, \bibinfo {author} {\bibfnamefont {Y.}~\bibnamefont
			{Zhang}}, \bibinfo {author} {\bibfnamefont {A.~C.}\ \bibnamefont {Komarek}},
		\bibinfo {author} {\bibfnamefont {E.}~\bibnamefont {Kampert}}, \emph
		{et~al.},\ }\bibfield  {title} {\bibinfo {title} {{Anomalous hall effect in
				weyl semimetal half-heusler compounds RPtBi (R= Gd and Nd)}},\ }\href
	{https://doi.org/https://doi.org/10.1073/pnas.1810842115} {\bibfield
		{journal} {\bibinfo  {journal} {PNAS}\ }\textbf {\bibinfo {volume} {115}},\
		\bibinfo {pages} {9140} (\bibinfo {year} {2018})}\BibitemShut {NoStop}%
	\bibitem [{\citenamefont {Zhu}\ \emph {et~al.}(2020{\natexlab{b}})\citenamefont
		{Zhu}, \citenamefont {Singh}, \citenamefont {Wang}, \citenamefont {Huang},
		\citenamefont {Chiu}, \citenamefont {Wang}, \citenamefont {Graf},
		\citenamefont {Zhang}, \citenamefont {Lin}, \citenamefont {Sun} \emph
		{et~al.}}]{zhu2020exceptionally}%
	\BibitemOpen
	\bibfield  {author} {\bibinfo {author} {\bibfnamefont {Y.}~\bibnamefont
			{Zhu}}, \bibinfo {author} {\bibfnamefont {B.}~\bibnamefont {Singh}}, \bibinfo
		{author} {\bibfnamefont {Y.}~\bibnamefont {Wang}}, \bibinfo {author}
		{\bibfnamefont {C.-Y.}\ \bibnamefont {Huang}}, \bibinfo {author}
		{\bibfnamefont {W.-C.}\ \bibnamefont {Chiu}}, \bibinfo {author}
		{\bibfnamefont {B.}~\bibnamefont {Wang}}, \bibinfo {author} {\bibfnamefont
			{D.}~\bibnamefont {Graf}}, \bibinfo {author} {\bibfnamefont {Y.}~\bibnamefont
			{Zhang}}, \bibinfo {author} {\bibfnamefont {H.}~\bibnamefont {Lin}}, \bibinfo
		{author} {\bibfnamefont {J.}~\bibnamefont {Sun}}, \emph {et~al.},\ }\bibfield
	{title} {\bibinfo {title} {{Exceptionally large anomalous Hall effect due to
				anticrossing of spin-split bands in the antiferromagnetic half-Heusler
				compound TbPtBi}},\ }\href
	{https://doi.org/https://doi.org/10.1103/PhysRevB.101.161105} {\bibfield
		{journal} {\bibinfo  {journal} {Physical Review B}\ }\textbf {\bibinfo
			{volume} {101}},\ \bibinfo {pages} {161105} (\bibinfo {year}
		{2020}{\natexlab{b}})}\BibitemShut {NoStop}%
	\bibitem [{\citenamefont {Suzuki}\ \emph {et~al.}(2016)\citenamefont {Suzuki},
		\citenamefont {Chisnell}, \citenamefont {Devarakonda}, \citenamefont {Liu},
		\citenamefont {Feng}, \citenamefont {Xiao}, \citenamefont {Lynn},\ and\
		\citenamefont {Checkelsky}}]{suzuki2016large}%
	\BibitemOpen
	\bibfield  {author} {\bibinfo {author} {\bibfnamefont {T.}~\bibnamefont
			{Suzuki}}, \bibinfo {author} {\bibfnamefont {R.}~\bibnamefont {Chisnell}},
		\bibinfo {author} {\bibfnamefont {A.}~\bibnamefont {Devarakonda}}, \bibinfo
		{author} {\bibfnamefont {Y.-T.}\ \bibnamefont {Liu}}, \bibinfo {author}
		{\bibfnamefont {W.}~\bibnamefont {Feng}}, \bibinfo {author} {\bibfnamefont
			{D.}~\bibnamefont {Xiao}}, \bibinfo {author} {\bibfnamefont {J.~W.}\
			\bibnamefont {Lynn}},\ and\ \bibinfo {author} {\bibfnamefont
			{J.}~\bibnamefont {Checkelsky}},\ }\bibfield  {title} {\bibinfo {title}
		{Large anomalous hall effect in a half-heusler antiferromagnet},\ }\href
	{https://www.nature.com/articles/nphys3831} {\bibfield  {journal} {\bibinfo
			{journal} {Nat. Phys.}\ }\textbf {\bibinfo {volume} {12}},\ \bibinfo {pages}
		{1119} (\bibinfo {year} {2016})}\BibitemShut {NoStop}%
	\bibitem [{\citenamefont {Yao}\ \emph {et~al.}(2004)\citenamefont {Yao},
		\citenamefont {Kleinman}, \citenamefont {MacDonald}, \citenamefont {Sinova},
		\citenamefont {Jungwirth}, \citenamefont {Wang}, \citenamefont {Wang},\ and\
		\citenamefont {Niu}}]{AHC}%
	\BibitemOpen
	\bibfield  {author} {\bibinfo {author} {\bibfnamefont {Y.}~\bibnamefont
			{Yao}}, \bibinfo {author} {\bibfnamefont {L.}~\bibnamefont {Kleinman}},
		\bibinfo {author} {\bibfnamefont {A.~H.}\ \bibnamefont {MacDonald}}, \bibinfo
		{author} {\bibfnamefont {J.}~\bibnamefont {Sinova}}, \bibinfo {author}
		{\bibfnamefont {T.}~\bibnamefont {Jungwirth}}, \bibinfo {author}
		{\bibfnamefont {D.-s.}\ \bibnamefont {Wang}}, \bibinfo {author}
		{\bibfnamefont {E.}~\bibnamefont {Wang}},\ and\ \bibinfo {author}
		{\bibfnamefont {Q.}~\bibnamefont {Niu}},\ }\bibfield  {title} {\bibinfo
		{title} {First principles calculation of anomalous hall conductivity in
			ferromagnetic bcc {Fe}},\ }\href
	{https://doi.org/10.1103/PhysRevLett.92.037204} {\bibfield  {journal}
		{\bibinfo  {journal} {Phys. Rev. Lett.}\ }\textbf {\bibinfo {volume} {92}},\
		\bibinfo {pages} {037204} (\bibinfo {year} {2004})}\BibitemShut {NoStop}%
\end{thebibliography}

%

\end{document}